\documentclass{aa}  
\usepackage[utf8]{inputenc}
\usepackage{makecell} 
\usepackage{graphicx}
\usepackage{txfonts}
\usepackage{lipsum}
\usepackage[]{hyperref}
\usepackage{subcaption}         

\usepackage{lscape}

\usepackage{placeins}

\usepackage{multirow}
\usepackage{orcidlink}

\usepackage{color}

\begin{document}

   \title{Exploring the surface of HD 189733 via Doppler shadow analysis of planetary transits}

    \author{E. C. Gonçalves\inst{\ref{inst1}, \ref{inst3}}\orcidlink{0009-0003-0218-5010},
            E. Cristo\inst{\ref{inst1}}\orcidlink{0000-0001-5992-7589},
            W. Dethier\inst{\ref{inst1}}\orcidlink{0009-0007-3622-0928},
            N. C. Santos\inst{\ref{inst1}, \ref{inst3}}\orcidlink{0000-0003-4422-2919},
            S. G. Sousa\inst{\ref{inst1}}\orcidlink{0000-0001-9047-2965},
            P. T. P. Viana\inst{\ref{inst1}, \ref{inst3}}\orcidlink{0000-0003-1572-8531},
            T. Azevedo Silva\inst{\ref{inst2}}\orcidlink{0000-0002-9379-4895},
            R. Allart\inst{\ref{inst4},\ref{inst5}}\orcidlink{0000-0002-1199-9759},
            V. Bourrier\inst{\ref{inst5}}\orcidlink{0000-0002-9148-034X}
            }

   \institute{ Instituto de Astrofísica e Ciências do Espaço, Universidade do Porto, CAUP, Rua das Estrelas, 4150-762 Porto, Portugal\label{inst1}
    \and
       Departamento de F\'{\i}sica e Astronomia, Faculdade de Ci\^encias, Universidade do Porto, Rua do Campo Alegre, 4169-007 Porto, Portugal\label{inst3}
    \and
       INAF – Osservatorio Astrofisico di Arcetri, Largo E. Fermi 5, 50125
    Firenze, Italy\label{inst2}
    \and
    D\'epartement de Physique, Institut Trottier de Recherche sur les Exoplan\`etes, Universit\'e de Montr\'eal, Montr\'eal, Qu\'ebec, H3T 1J4, Canada\label{inst4}
    \and
    Observatoire Astronomique de l'Universit\'e de Gen\`eve, Chemin Pegasi 51, Sauverny, CH-1290, Switzerland\label{inst5}
             }

   \date{Received February 17, 2026}

       \titlerunning{Exploring the surface of HD 189733 via Doppler shadow analysis of planetary transits}
   \authorrunning{E. C. Gonçalves et al.}

  \abstract
   {Transmission spectroscopy has advanced our understanding of exoplanet atmospheres, but it can be hindered by contamination originating in stellar heterogeneities, mainly coming from line-of-sight effects. Therefore, probing how stellar spectra vary across the stellar surface is essential to accurately disentangling stellar and planetary spectral contributions in transit observations. Such observations can actually be used to reconstruct the local stellar spectra behind the planet's transit chord. These methodologies can help us learn more about the physics of stellar surface and how to tackle line-of-sight effects.}
   {In this paper, we study the centre-to-limb variations of line profiles across the surface of HD 189733 using the ESPRESSO spectrograph. We build on other works by analysing the same sets of \ion{Fe}{I} lines, allowing for a direct comparison of results and an assessment of the feasibility of applying the Doppler shadow technique with ESPRESSO. We gain a better understanding of the variations in line profiles, while also making a comparison between the data of HD 189733 and synthetic spectra and solar data.} 
   {We analysed spectra collected by ESPRESSO during two transits of HD 189733 b as separate sets of data. We performed a cross-correlation of each individual spectrum with two different masks made of selected \ion{Fe}{I} spectral lines for a total of four sets of cross-correlation functions (CCFs) and employed a Doppler shadow methodology to retrieve profiles for local regions of the stellar surface. We then compared the results with previous works and with solar disc-resolved observations from IAG ATLAS. Finally, we compared the data with two separate transit simulations made using SOAPv4 with Turbospectrum synthetic spectra computed with MARCS stellar atmosphere models under  local thermal equilibrium (LTE) and non-LTE (NLTE) conditions.}
   {For the profile depth of three sets of \ion{Fe}{I} CCFs, we verified a statistically significant increase in line depth from the stellar limb to the centre. This variation was expected from simulations with MARCS models, although the solar data present a smaller gradient in the variation of line depth. In the case of the width of the line profiles, we verified  that the profile width decreases from stellar limb to stellar centre for a set
of \ion{Fe}{I} CCFs. This result is consistent with the behaviour observed in solar data, but not reproduced by the simulations.}
   {These results highlight the abilities of ESPRESSO in providing the necessary precision and resolution to study centre-to-limb variations of spectral line profiles on the surface of other stars with the use of CCFs. The local CCF profiles of HD 189733 agree with the IAG ATLAS data, but disagree with simulations on line widths, indicating that important physical processes are missing and must be included to recover accurate profile widths.}


   \maketitle

\nolinenumbers
\section{Introduction}

The study of exoplanetary atmospheres provides invaluable information to supplement our understanding of their composition, structure, and formation. Despite the challenges involved, the number of planets with detected atmospheric signatures is rapidly increasing.
Recent examples of this success include the identification of a myriad of chemical species \citep[e.g.][]{Snellen_2010,Sing_2016,Azevedo_Silva_2022,Ehrenreich_2020,Madhusudhan_2023}, the characterisation of dynamical phenomena, such as winds \citep{Seidel_2025} using low-, medium-, and high-resolution spectroscopy, along with the measurement of atmospheric albedos and phase curves using high-precision photometry \citep{Demangeon_2024}.

Much of the progress in the field is based on the use of transmission (or transit) spectroscopy. This technique first succeeded with the detection of sodium in the atmosphere of the hot-Jupiter planet HD 209458 b \citep{Charbonneau_2002}, who identified a deeper transit in the sodium doublet by comparing in-transit and out-of-transit observations across different bandpasses. However, the classical approach used to extract the transmission spectrum assumes that the stellar spectrum behind the planet's atmosphere is well approximated by the out-of-transit spectrum. As shown, for example, in  \citet{Dethier_2023}, this is often far from being a good approximation, due to the combination of centre-to-limb variations (CLV) from the stellar surface and line-of-sight effects, which depend on the architecture of the star-planet system, leading to biased retrieval of atmospheric signals \citep[see also][]{Rackham_2018}. This situation is well illustrated by the doubts raised about the existence of sodium in HD 209458 b as shown in \citet{Casasayas-Barris_2020,Casasayas-Barris_2021}, where they demonstrated that the signals identified through transmission spectroscopy could be impacted by CLV and stellar rotational effects. 

To avoid this problem, it is fundamental to understand how stellar spectra vary across the stellar disc. While stellar atmosphere models, in principle, might be of assistance \citep[e.g.][]{Dravins_2024}, gauging their quality implies that we need to obtain disc-resolved observations of stars of different spectral types. However, transit spectroscopy time series can also be used to probe the disc of the host star. In fact, by subtracting the acquired spectra during the transit from the spectra obtained before or after the transit, we can (in principle) retrieve the spectra of the star hidden by the planet \citep{Cameron_2010} \footnote{Note: some signal is to be expected from the planet's atmosphere, but it is minor compared to the flux of the obscured region.}.

This approach was used to probe the stellar disc behind the planets HD 189733 b \citep{Cegla_2016,Dravins_2018}, HD 209458 b \citep{Dravins_2017}, and WASP-166 b \citep{Doyle_2022}, as well as WASP-52 b and HAT-P-30 b \citep{Cegla_2023}.
One of the biggest challenges of this approach is the low signal-to-noise ratio (S/N) of the retrieved data. Since the planet area is only (at maximum) a few percent of the stellar disc, the retrieved spectrum will typically have an  S/N  value that is at least one order of magnitude lower than that of the total spectrum of the star. As such, different studies have focused on using this method, together with cross-correlation techniques, to explore the behaviour of different groups of lines as a function of disc position. The results published so far are encouraging, but the S/N of the original data did not allow for the retrieval of the local stellar spectra to the necessary levels of precision to distinguish certain trends of centre-to-limb variations of line profiles,  as seen in the work by \citep{Dravins_2018}.

In this paper, we make use of the ESPRESSO high-resolution and high-S/N spectra of HD 189733, obtained during two different transits of HD 189733 b \citep{Bouchy_2005}, to analyse the change of the profiles of two sets of spectral \ion{Fe}{I} lines as a function of disc position. Our methodology and analysis build on the work and results of \citet{Dravins_2018}, using the HARPS spectrograph to study the centre-to-limb variations in HD 189733 of two specific sets of \ion{Fe}{I} lines. We  used the same sets of \ion{Fe}{I} lines to show improvements thanks to the availability of higher quality datasets obtained from ESPRESSO.
In Sect. 2, we introduce the datasets, in Sect. 3 we describe the Doppler shadow methodology employed, in Sect. 4 we present the results and offer comparisons to \citet{Dravins_2018} and model simulations, and in Sect. 5 we present our conclusions.

\section{The dataset}

The spectra were collected using the ESPRESSO spectrograph (ESO, Paranal) in the standard high-resolution mode ($\lambda/\Delta\lambda \sim$140\,000), covering the full optical domain from 380 to 780 nm. The two sets of observations studied in this paper were taken using the UT1 telescope at the VLT. The first was on 11 August 2021, between 00:45 and 04:45 UTC, when 41 exposures were made, and the second on 31 August 2021, between 00:27 and 04:40 UTC, when 43 exposures were made. The exposure times were fixed at 300 seconds for both nights. This resulted in an average S/N of 159 and 144, around 570 nm, for the first and second nights, respectively. These same observations were first used in \citet{Cristo2024} to study the broadband transmission spectrum of HD 189733 b. We point to \citet{Cristo2024} for a more thorough description of the observations. The last observation on 11 August 2021 had a significantly lower S/N (i.e. 39) and, thus, it was discarded.

\begin{figure}
   \centering
   \includegraphics[width=\hsize]{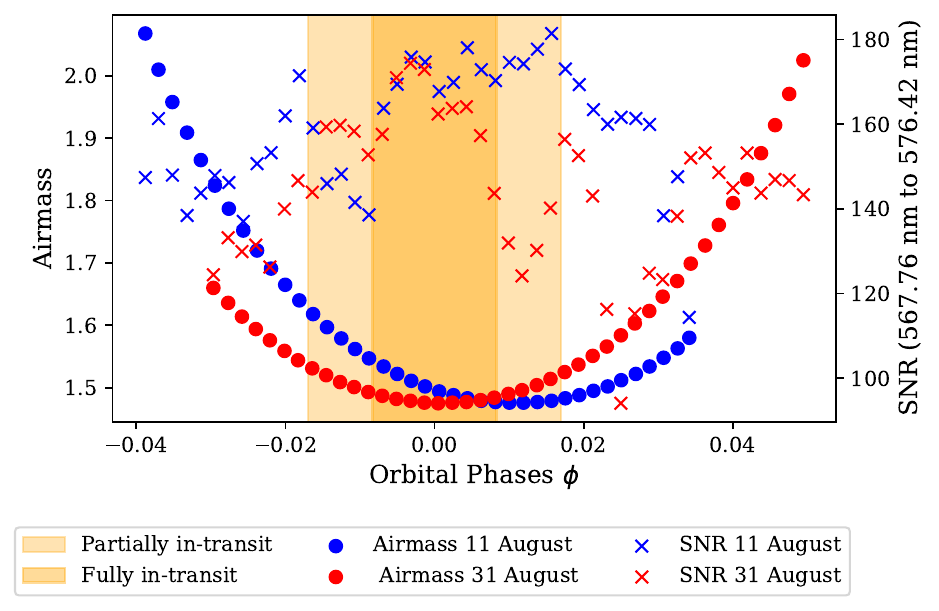}
    \caption{Airmass variation and S/N of an ESPRESSO spectral order covering wavelengths from 567.76 nm to  576.42 nm, chosen as a midpoint in the selected \ion{Fe}{I} spectral lines, as a function of the orbital phase of HD 189733 b. Blue points correspond to observations from 11 August 2021, and red points to those from 31 August 2021. The darker shaded region marks the phases where the planet is fully in transit, with lighter shading indicating the ingress and egress phases.}
    \label{fig:t_obs}
\end{figure}

An overview of the observations relative to the orbital phases of the planet is given in Fig. \ref{fig:t_obs}. On each of the two nights, there were 18 observations acquired during the transit. The data were reduced using the ESPRESSO pipeline, DRS version 3.0.0 \citep[for more details see][]{Cristo2024}. All analyses in this paper were performed using 1D spectra. In addition, all spectra used in this work were already included in the reference frame of the barycentre of the HD 189733 system.

\section{Methodology}
\subsection{Telluric correction}

One of the big obstacles for any high-precision spectroscopic study using ground-based data is the absorption lines from Earth's atmosphere (primarily oxygen and water vapour).  To remove these from the regions of interest, we decided to use the ESO code \texttt{Molecfit} \citep{Smette_2015,Kausch_2015} version 4.3.1. \texttt{Molecfit} is a tool used to remove absorption features from a spectrum by fitting synthetic Earth transmission spectra with real data. 

When adjusting the settings for the telluric correction, we used the fitting parameters as in \citet{Azevedo_Silva_2022}. These values were originally used to clean from tellurics the data of WASP-76 b and WASP-121 b, which were also taken with ESPRESSO. The parameters for the telluric correction were kept constant for the full time-series of stellar spectra.

\begin{figure}
    \centering
    \includegraphics[width=\linewidth]{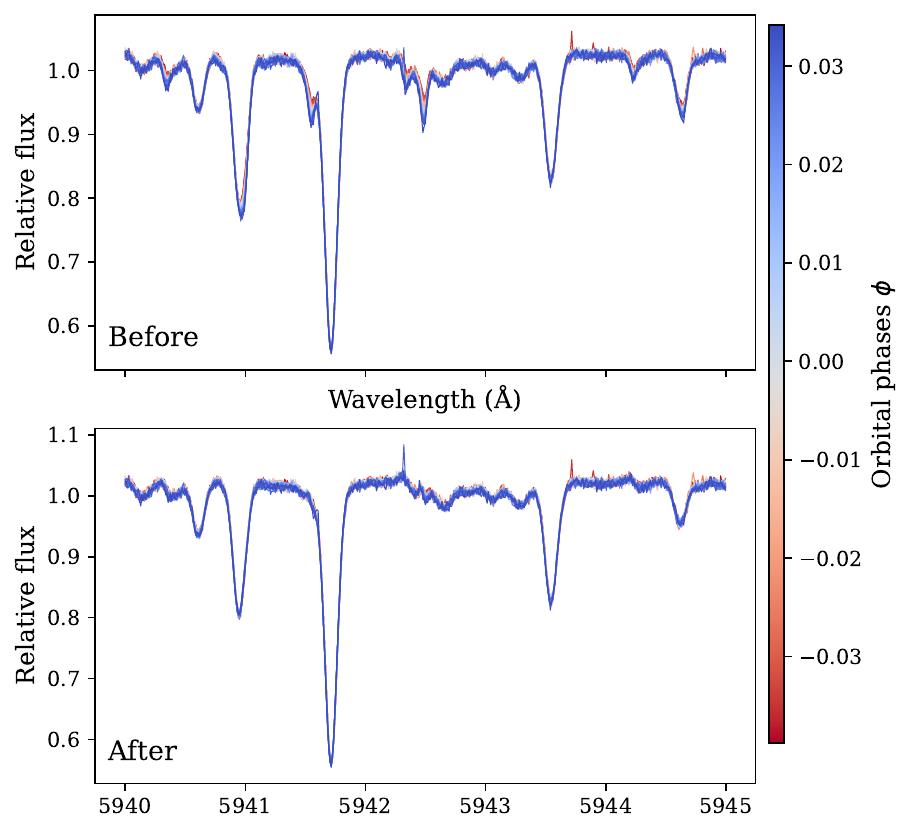}
    \caption{Comparison of the same section of the spectra (normalised to continuum) before (top) and after telluric correction with Molecfit (bottom). The colour bar represents the orbital phases of the corresponding spectra. The telluric lines can be seen varying in depth due to changes in airmass before the correction.}
    \label{fig:telluric}
\end{figure}

In Fig. \ref{fig:telluric}, we see the difference for a small wavelength range of the spectra for the night of observations on 11 August 2021 before and after the correction. For the uncorrected data, we see lines whose depths vary over the course of the transit, as they are colour-marked according to their orbital phase. This is a typical behaviour of Earth's atmospheric lines, where the amount of absorption is dependent (among other aspects) on the airmass along our line of sight, which varies over the course of the night of observations, lasting for about 4 hours (Fig. \ref{fig:t_obs}). In contrast, the corrected spectra have the telluric lines essentially removed, resulting in spectra whose lines are independent of changes in airmass.
By looking at the ratio of the uncorrected over the telluric-corrected spectra shown in Figs. \ref{fig:t_left} and \ref{fig:t_right}, we can observe in the top panels a clear variation in the impact of tellurics on flux for a specific wavelength interval (the same as Fig. \ref{fig:telluric}). The middle panels of Figs. \ref{fig:t_left} and \ref{fig:t_right} span the entire wavelength domain of ESPRESSO and we can see that for wavelengths greater than 690 nm, the telluric lines become oversaturated and, thus, the data are not reliable. Lastly, the bottom panels show the mean flux of the fitted tellurics plotted with airmass,  visually showcasing  a great correlation between the two and indicating that the telluric correction was successful, as this behaviour was indeed expected.

\subsection{Cross-correlation}

\begin{figure}
    \centering
    \includegraphics[width=\linewidth]{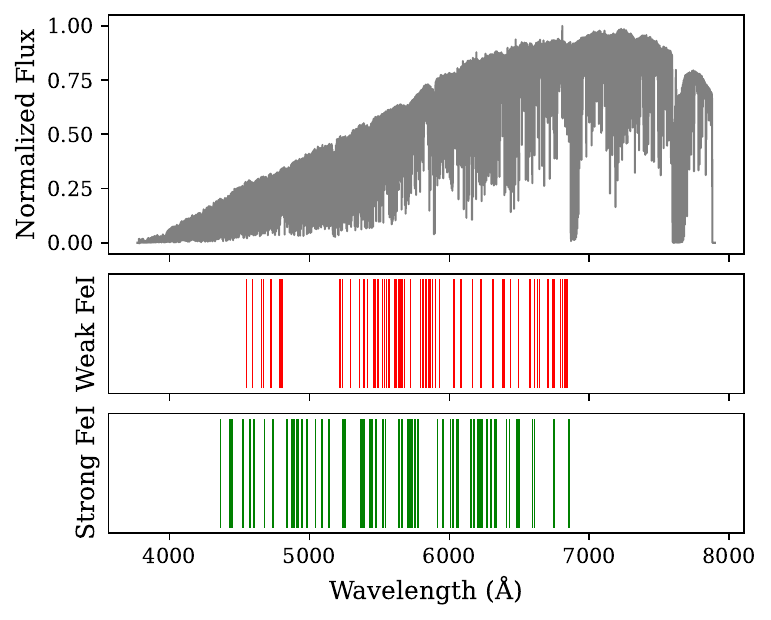}
    \caption{Spectral line positions of \ion{Fe}{I} lines taken from \citet{Dravins_2018} with one example spectrum (uncorrected for tellurics). Top: Example spectrum shown in grey. Middle: Wavelength positions of the weak \ion{Fe}{I} lines shown as red vertical lines. Bottom:  Wavelength positions of the strong \ion{Fe}{I} lines, shown as green vertical lines.}
    \label{fig:ironlines}
\end{figure}

To work with higher S/N data, we performed a cross-correlation of the spectra with masks made for specific spectral lines to create cross-correlation functions (CCFs). This approach is often used in transmission spectroscopy measurements \citep[e.g.][]{Ehrenreich_2020}.
The chosen \ion{Fe}{I} lines were taken from \citet[][]{Dravins_2018}. In their work, 158 \ion{Fe}{I} lines were carefully selected to avoid blended lines \citep[Tables E.1 and E.2 of][]{Dravins_2018}. Their high formation temperature, compared to that of the atmosphere of the planet, allows us to treat the planet as an opaque body with a transparent atmosphere. They also separated these lines into two groups, strong and weak \ion{Fe}{I} lines, based on their absorption depth in relation to the local continuum, with 55\% being the division criterion. These two main groups were used to construct the masks used for cross-correlation. The location of their central wavelength in relation to the observed spectra is shown in Figure \ref{fig:ironlines}.

There are 86 weak \ion{Fe}{I} lines and they range in wavelength position from 455 nm to 685 nm, their excitation potential ranging from 0.99 eV to 4.95 eV. There are then 72 strong \ion{Fe}{I} lines and they range from 436 nm to 686 nm, their excitation potential ranging from 0.09 eV to 4.73 eV \citep[][]{Dravins_2018}. These lines were selected for a study of HD 189733 using HARPS data, which has a smaller wavelength domain than ESPRESSO, but it is still useful for our purposes; especially as it stops short of the telluric oversaturated oxygen B band region beyond 686 nm.

From the list of weak and strong \ion{Fe}{I} lines, simple binary masks were created, using an approach similar to the one used, for example, for the ESPRESSO spectrograph \citep[][]{Pepe_2021}. Each binary aperture has a width of 500 $\rm m\,s^{-1}$ (the resolution element of ESPRESSO).

To create the CCFs, we used the \texttt{iCCF} Python package \footnote{iCCF PyPI page: https://pypi.org/project/iCCF/} on the telluric-corrected spectra. With these two masks, we cross-correlated all telluric-corrected spectra. This resulted in two types of CCFs for each of the two days of observation. These CCFs were used for all subsequent analyses.

\subsection{Doppler shadow}

The Doppler shadow method was first introduced in \citet{Cameron_2010} and was successfully explored in several subsequent papers \citep[e.g.][]{Cegla_2016,Dravins_2017_I,Dravins_2017,Dravins_2018}. This technique allows us to retrieve spectral information from local regions of the stellar surface that are in the 'shadow' of the transiting planet. By probing the local spectra of a star, this method offers the possibility of studying the behaviour of different stellar lines with origin at different positions in the stellar disc. We point to the above-mentioned papers for details on the method. These concepts have also been included in the ANTARESS workflow \citep{bourrier_2024} to extract local stellar spectra that were used to study HD 209458 b and WASP-76 b with ESPRESSO data.

\begin{figure}
    \centering
    \includegraphics[width=1.\linewidth]{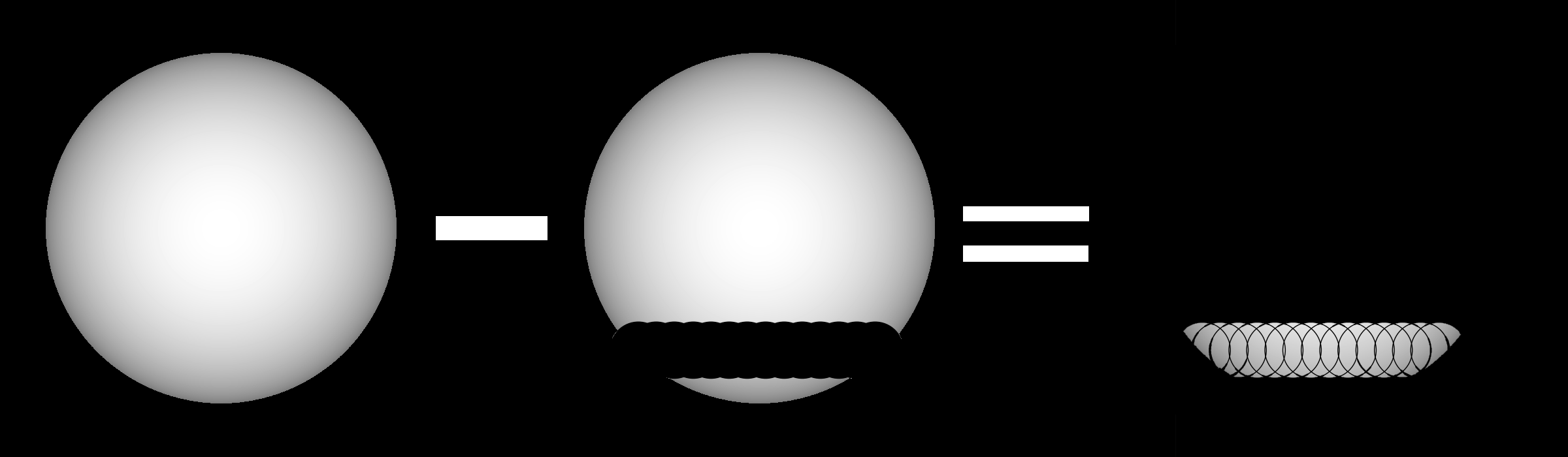}
    \caption{Schematic of the Doppler shadow method for the system HD 189733 b.}
    \label{fig:subtract}
\end{figure}

The principle behind the Doppler shadow method is straightforward. For a star with a transiting planet, the planet will obscure successive parts of the stellar disc as the transit progresses. By subtracting the in-transit observations from the out-of-transit observations, we retrieved the information about the covered regions for each exposure, that is, for successive regions of the stellar surface along the transit chord. Depending on the orbital architecture of the system, this technique can potentially allow us to scan different stellar latitudes. In Fig. \ref{fig:subtract}, we present a diagram of the idea using the system HD 189733 studied in this work as an illustration of this approach.

To retrieve the Doppler shadow spectra (henceforth referred to as the local CCFs) from our dataset, we first needed to normalise the CCFs and correct the RVs to the stellar rest frame. We started by fitting the disc-integrated CCFs to measure the radial velocities at each given epoch. We decided to use the modified Gaussian function presented in \citet{Dravins_2017_I,Dravins_2017,Dravins_2018} to remain consistent with them and to compare the results later. They argued that these functions provide better fits than, for example, Lorentzian profiles. The function is
\begin{equation}
    \text{f}\left(x\right)=y_{0}+a\exp\left(-\frac{1}{2}\left(\frac{\left|x-x_{0}\right|}{b}\right)^{c}\right)
    \label{eq:2.1}
,\end{equation}

where $y_0$ is the value of the continuum and will be used to normalise the CCFs, $x_0$ is the central radial velocity of the CCF, $a$ is a scaling parameter, $b$ is equivalent to the variance of a regular Gaussian function that changes the profile width, and $c$ is a shape parameter (exponent) controlling the departure from a Gaussian profile. The case $c=2$ corresponds to a standard Gaussian, whereas $c<2$ yields profiles with sharper core and heavier tails and $c>2$ produces flatter-bottomed profiles with lighter tails.

To perform the profile fitting, we used the $curve\_fit$ function of the \texttt{scipy} Python package \citep[$curve\_fit$ from][]{2020SciPy-NMeth,Vugrin_2007}. This routine returns the covariance matrix of the fitting parameters from which we took the square root of the diagonal as our uncertainties (i.e. standard deviations).

Before correcting for the barycentric motion of the system by fitting the star's velocity, we first identify which observations are in-transit and which are out-of-transit, using the ephemerides presented in Table \ref{tab:1}. In-transit observations are defined as those occurring within half a transit duration of the transit epoch ($\rm T_0$), and the transit duration ($\Delta T$) is calculated with the simplified formula of \citet{Seager_2003}. To perform this comparison, we converted the observation times and transit duration into orbital phases ($\phi$) and then converted the transit epoch ($\rm T_0$) from MBJD to BJD. Because $\rm T_0$ was measured in 2009, we also applied a correction by subtracting the respective night's $\Delta \phi_{\rm p,0}$ and $\Delta \phi_{\rm p,1}$ values multiplied by the orbital period (see Table \ref{tab:1}).

\begin{figure}[h!]
   \centering
   \includegraphics[width=\hsize]{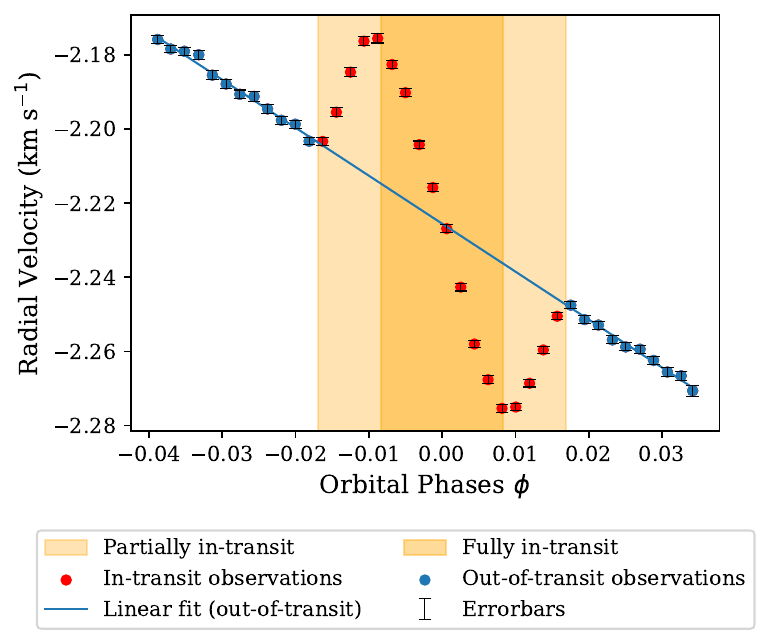}
      \caption{Radial velocities of the strong \ion{Fe}{I} CCFs for the 11 of August 2021. The shaded area represents the time of transit, with the lighter sections representing the ingress and egress of the transit. The dots in red represent the in-transit observations, while blue represent out-of-transit observations. The blue line represents the linear fit to the motion of the star around its system's barycentre.}
         \label{fig:bary_strongfe}
\end{figure}

\begin{table}
    \begin{center}
    \caption{HD 189733's system parameters used in this work.}
    \begin{tabular}{c c c}
        \hline\hline
        Parameter & Value & References \\
        \hline
        $\rm P_{rot}$ (days) & $11.454_{-0.088}^{+0.092}$ & Cr24 \\
        $\rm i_{star}\,(^{\circ})$ & $71.87_{-6.91}^{+5.55}$ & Cr24 \\
        $\rm i_{planet}\,(^{\circ})$ & $85.465_{-0.021}^{+0.020}$ & Cr24 \\
        $\lambda \,(^{\circ})$ & $-1.00_{-0.23}^{+0.22}$ & Cr24 \\
        $\rm \rm a_{planet}\left(R_{{star}}\right)$ & $8.7686_{-0.0082}^{+0.0082}$ & Cr24\\
        $\rm R_{planet}\left(R_{star}\right)$ & $0.1602_{-0.0035}^{+0.0039}$ & Cr24 \\
        $\Delta\phi_{\rm p,0}$ & $-0.002424_{-0.00035}^{+0.00036}$ & Cr24 \\
        $\Delta\phi_{\rm p,1}$ & $-0.002300_{-0.00044}^{+0.00035}$ & Cr24 \\
        $\rm R_{star} \,(R_\odot)$ & $0.766_{-0.013}^{+0.007}$ & Tr09 \\
        $\rm T_{eff} $ (K) & $4969 \pm 43$ & So18\\
        $\text{T}_{0}$ (MBJD) & $53988.30339_{-0.000039}^{+0.000072}$ & Tr09\\
        $\rm P_{orbit}$ (days) & $2.21857312^{+0.00000036}_{-0.00000076}$ & Tr09 \\ 
        $\rm \left[Fe/H\right]$ (dex) & $-0.07 \pm 0.02$ & So18 \\
        $\rm \log\left(g\right)$ & 4.60 $\pm 0.01$ & So18 \\
        \hline
    \end{tabular}
    \tablefoot{$\rm P_{rot}$ is the rotation period of the HD 189733, $\rm i_{star}$ is the stellar inclination, $\rm i_{planet}$ is the orbital inclination of the planet, $\lambda$ is the spin-orbit angle, $\rm a_{planet}\left(R_{{star}}\right)$ is the semi-major axis in units of the stellar radius, $\rm R_{planet}\left(R_{star}\right)$ is the planetary radius in units of the stellar radius. $\Delta\phi_{\rm p,0}$ and $\Delta\phi_{\rm p,1}$ are dimensionless correction factors for the respective transit epochs of the transits of 11 August 2021 and 31 August 2021. $\rm R_{star}$ is the stellar radius, $\rm T_{eff} $ is the effective temperature, $\text{T}_{0}$ is the transit epoch of the planet, $\rm P_{orbit}$ is the orbital period of the planet, $\rm \left[Fe/H\right]$ is the stellar metalicity, and $\rm \log\left(g\right)$ is the surface gravity of HD 189733.}
    \tablebib{(Cr24)~\citet{Cristo2024}; (Tr09) \citet{Triaud_2009}; (So18) \citet{Sousa_2018}}
    \label{tab:1}
    \end{center}
\end{table}

Although the whole out-of-transit motion of the star follows a Keplerian curve, given that the timeframe of the observations is much smaller than the orbital period, the motion can be approximated by a linear function. This can be seen in Fig. \ref{fig:bary_strongfe} for the case of the CCF RVs computed using the strong \ion{Fe}{I} line list.
Therefore, we used a linear regression to fit only the radial velocity points corresponding to the out-of-transit exposures. This allowed us to extract the RV component due to the star's motion around the barycentre, excluding the  component attributed to the Rossiter-McLaughlin (RM) effect. To apply the RV correction, we shifted each CCF by the RV value derived from the fit. The CCFs are then placed into the star's rest frame and only the RM effect remains seen in Fig. \ref{fig:bary_strongfe} prior to stellar rest frame correction. An example of the before and after of RV correction is shown for one set of CCFs built from the strong \ion{Fe}{I} lines in Fig. \ref{fig:bary_strongfe} (before) and left side of Fig. \ref{fig:appendix_bary_strongfe_2} (after), this process is identical for the other 3 RV corrections performed. The linear fits of the strong \ion{Fe}{I} case are within the error of the weak \ion{Fe}{I} case. Figures \ref{fig:appendix_bary_strongfe_2} and \ref{fig:appendix_bary_weakfe} show the result of the RV values corrected for the motion of the star around the barycentre for the set of strong and weak \ion{Fe}{I} lines CCFs, respectively.

\begin{figure}[h!]
   \centering
   \includegraphics[width=\hsize]{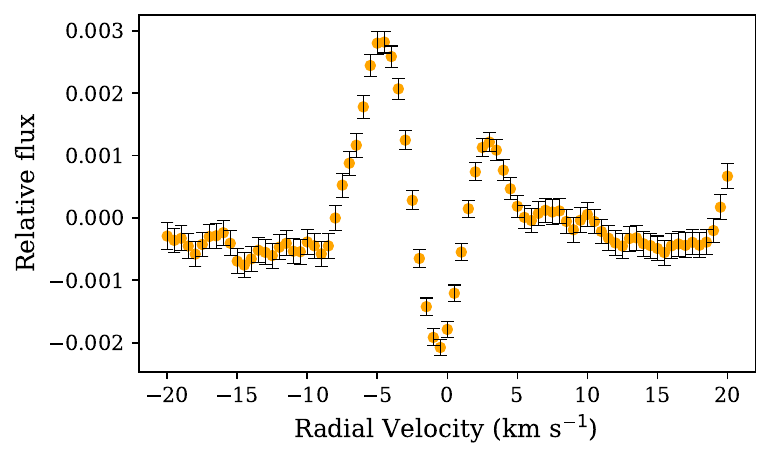}
   \includegraphics[width=\hsize]{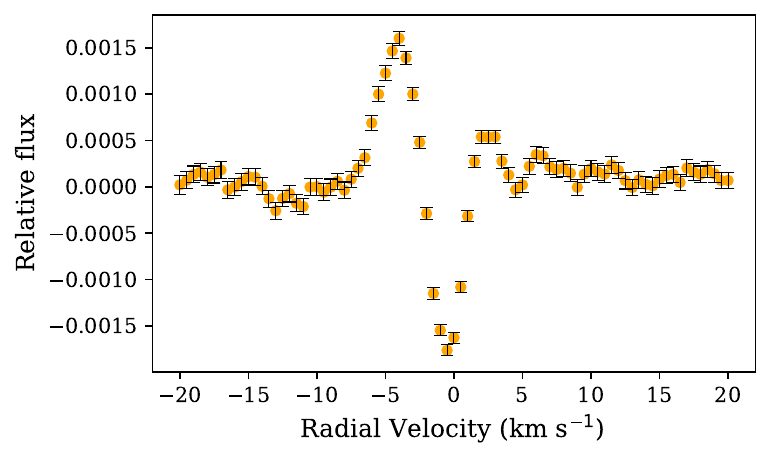}
      \caption{Difference in master out-of-transit of the two nights of observation (31 August subtracted by 11 August). Top: Case of the strong \ion{Fe}{I} lines CCFs. Bottom: Case of the weak \ion{Fe}{I} lines CCFs.}
         \label{fig:master_dif}
\end{figure}

With the CCFs in the stellar rest frame, we proceeded to apply the Doppler shadow method, which requires the construction of a master out-of-transit CCF and correction of the normalised flux before subtracting. For each night, we averaged all RV-corrected out-of-transit CCFs to produce the master out-of-transit CCF. We compared the master CCFs of the first and second nights by subtracting the latter by the former to see if there were any changes, as there is a separation of 20 days between sets of observations. This was done for the CCFs of the weak and strong \ion{Fe}{I} lines as shown in Fig. \ref{fig:master_dif}. 
A clear difference is visible between the two master out-of-transit CCFs, likely reflecting a change in the activity level of HD 189733 as the presence of active regions induces distortions in line profiles and other contributions that result in the observed shapes. To investigate this further, we used the \texttt{ACTIN 2} Python package \citep{Silva_2018,Silva_2021} to compute \ion{Ca}{II} activity indicators for the two transit epochs. For this, we used the average indicator value computed from the out-of-transit observations, which produced a value of $0.4616 \pm 0.0002$ for the first night and $0.4479 \pm 0.0002$ for the second night. These measurements support the interpretation that the two nights exhibited different stellar activity levels, with the first night being more active.

The in-transit observations needed to be scaled to their appropriate relative flux in comparison with the out-of-transit observations. This was done by using the photometric transit light curve of HD 189733 b. To build the light curve, we used version 4.0 of the Spot Oscillations And Planet (\texttt{SOAPv4}) code \citep{Cristo_2025}, which simulates the transit of a planet in front of a stellar surface with optional star spots.
Using the simulated light curve, we multiplied each RV-corrected in-transit CCF by the corresponding flux as in Eq. \ref{eq:2.sub}.

The \texttt{SOAPv4} light curve simulation requires values for the planet's orbital period ($\rm P_{orbit}$), eccentricity ($\rm e$), which we fixed at 0 since the orbit has known low eccentricity, orbital inclination ($\rm i_{planet}$), spin-orbit angle ($\lambda$), semi-major axis ($\rm a_{planet}$), and radius ($\rm R_{planet}$);  in addtion, for the star, we required its rotational period ($\rm P_{ rot}$), stellar inclination ($\rm i_{star}$), and radius ($\rm R_{star}$), all given in Table \ref{tab:1}. In addition, the \texttt{SOAPv4} simulation also requires the quadratic limb darkening coefficients. For this we used the \texttt{LDTK} Python package \citep{Parviainen2015,Husser2013}, which calculates the coefficients from the star's effective temperature ($\rm T_{eff}$), metallicity ($\rm \left[Fe/H\right]$), and surface gravity ($\rm \log\left(g\right)$) also found in Table \ref{tab:1}. We then selected the range of wavelengths of the ESPRESSO spectrograph. Finally, to extract the Doppler shadow, we took the series of scaled in-transit CCFs and subtracted them from the master out-of-transit CCF using

\begin{equation}
    F_{\rm local}(t)=f_{\rm out}-f_{\rm in}(t)\times F_{\rm SOAP}(t)
    \label{eq:2.sub}
,\end{equation}

where $f_{\text{out}}$ is the normalised master out-of-transit CCF, $ f_{\text{in}}$ are the normalised in-transit CCFs, and $F_{\text{SOAP}}$ is the simulated photometric flux given by \texttt{SOAPv4}. Here, $F_{\text{local}}$ is the Doppler shadow CCF flux or the local CCFs from the respective areas of the stellar surface occulted by the planet at each exposure during its transit at observation time, $t$.

\begin{figure}[h!]
   \centering
   \includegraphics[width=\hsize]{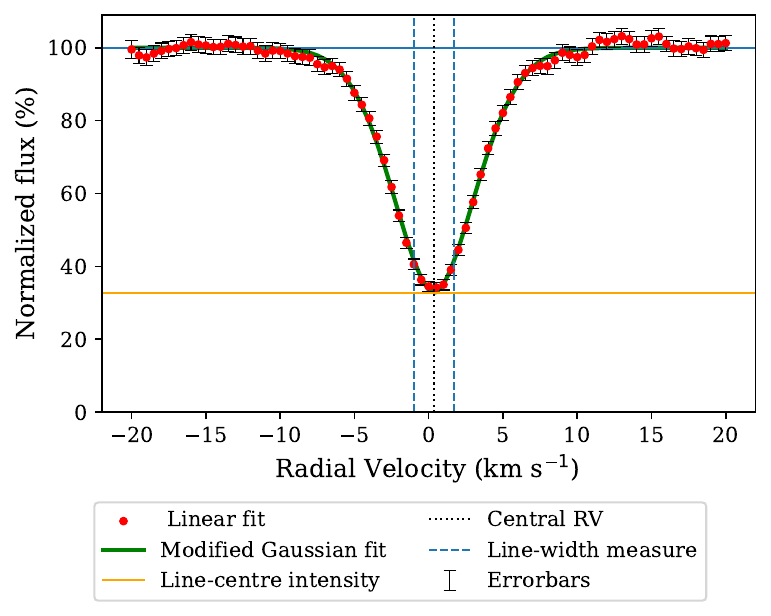}
      \caption{Example of a local CCF profile being fitted to a modified Gaussian function (see Eq. \ref{eq:2.1}). This specific example is one of the strong \ion{Fe}{I} local CCFs at $\phi=0.0157$ on the first night. This fit gives parameters with values $y_0=0.02741\pm0.0001$, $a=0.0185\pm0.0002$, $x_0 = 0.36\pm0.03$ km/s, $b=2.71\pm0.06$ km/s, and $c=1.81\pm0.07$.
      The red dots represent the measurements of the local CCF profile, the green solid line is the modified Gaussian fit, the horizontal blue line denotes the continuum level, the yellow horizontal line is the value of the derived line-centre intensity, the vertical black dotted line is the value of the central radial velocity, and the distance between the two dashed vertical blue lines represents the line-width measure.}
         \label{fig:fit_example}
\end{figure}

With this information, we can evaluate how the local CCF profiles change over the transit chord on the star's surface; namely, how the stellar spectrum changes locally over the surface of the star.  To derive information about these profile variations, we fit the local CCFs with a modified Gaussian function (Eq. \ref{eq:2.1}) to extract the values of $y_0$, $x_0$, $a$, and $b$. These parameters were used to extract the following quantities, namely, the central radial velocity ($x_0$), the line width ($b$), and the line-centre intensity ($I$) given by

\begin{equation}
    I=\left(1-\frac{a}{y_0}\right)\times100
    \label{eq:sus}
,\end{equation}

as defined in \citet{Dravins_2018}. A schematic of how these quantities describe the CCF profiles is given in Fig. \ref{fig:fit_example}. We also fit the master out-of-transit CCF to get reference values from the disc-integrated CCF.

Although orbital phases, $\phi$, are optimal for plotting quantities by observations time, it is useful to plot extracted quantities in relation to their position in the stellar disc from the disc centre. For this purpose, we converted $\phi$ into $\mu$, which is defined as the cosine of the angle between our line of sight and the normal to the stellar surface.

In principle, we could use a simple geometric model, where $\mu$  is calculated from the position of the planet’s disc centre at each observation. However, this approach can bias $\mu$ towards lower values near the limb due to limb darkening. To obtain a value of $\mu$ that is representative of the region of the star occulted by the planet, we chose to adopt a brightness-weighted approach similar to that of \citet{Cegla_2016}. We divided the stellar disc into a 600 by 600 square grid to obtain convergence and for each point in the grid calculated the local value of $\rm \mu_{local}$ and used the same quadratic limb darkening coefficients calculated with \texttt{LDTK} to calculate the local brightness for each point in the grid. Then, for each observation, we considered only the contributions of the grid of the region occulted of the planet to calculate a brightness-weighed value of $\mu$.

We discarded observations with $\mu$ less than 0.3 and their data were not included in Figs. \ref{fig:results_weakfei_ccf} and \ref{fig:results_strongfei_ccf}. These points correspond to limb regions, where the flux difference between the in-transit observations near the limb and the out-of-transit CCF is significantly smaller compared to the centre. This small difference leads to noisy local CCFs after subtraction. The local CCFs of these discarded limb regions after fitting have uncertainties at least an order of magnitude larger than the remaining local CCFs. For the multiple steps performed in our methodology, we give the propagation of the uncertainties in Appendix \ref{Appendix_C}.

\section{Results}

\subsection{Analysis}

\begin{figure*}[h]
    \centering
    \includegraphics[width=\linewidth]{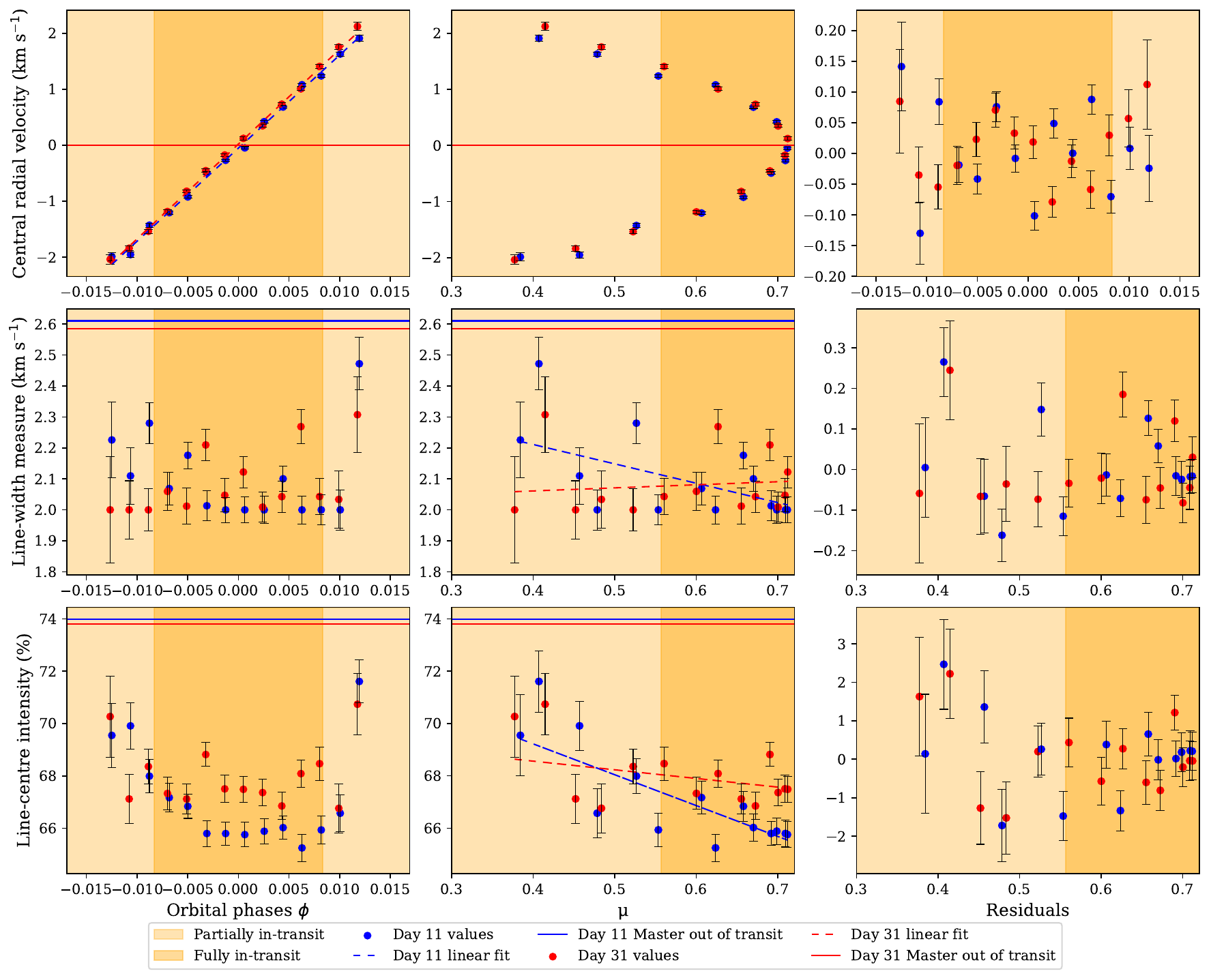}
    \caption{Datasets extracted from the local weak \ion{Fe}{I} lines CCFs via function fitting, where the three quantities shown are represented in Fig. \ref{fig:fit_example}. The dots show the data points, the solid lines represent the values of the master out-of-transit and the dashed lines are linear fits to the data serving only as a visual guide. Features coloured in blue refer to data from the 11 August 2021, while red shows data from the 31 August 2021. Lastly the shaded area represents the time of transit with the lightly coloured region being the ingress and egress of the transit.
    The top row of panels shows the data for the central radial velocities of the local CCFs, the middle row shows the data regarding to the line-width measure and the bottom row shows the data for the line-centre intensity. The left column shows the multiple data points as a function of the respective orbital phases ($\phi$), the middle column as a function of $\mu$, and the right column shows the residuals of the linear fits performed.}
    \label{fig:results_weakfei_ccf}
\end{figure*}

\begin{figure*}[h]
    \centering
    \includegraphics[width=1.\linewidth]{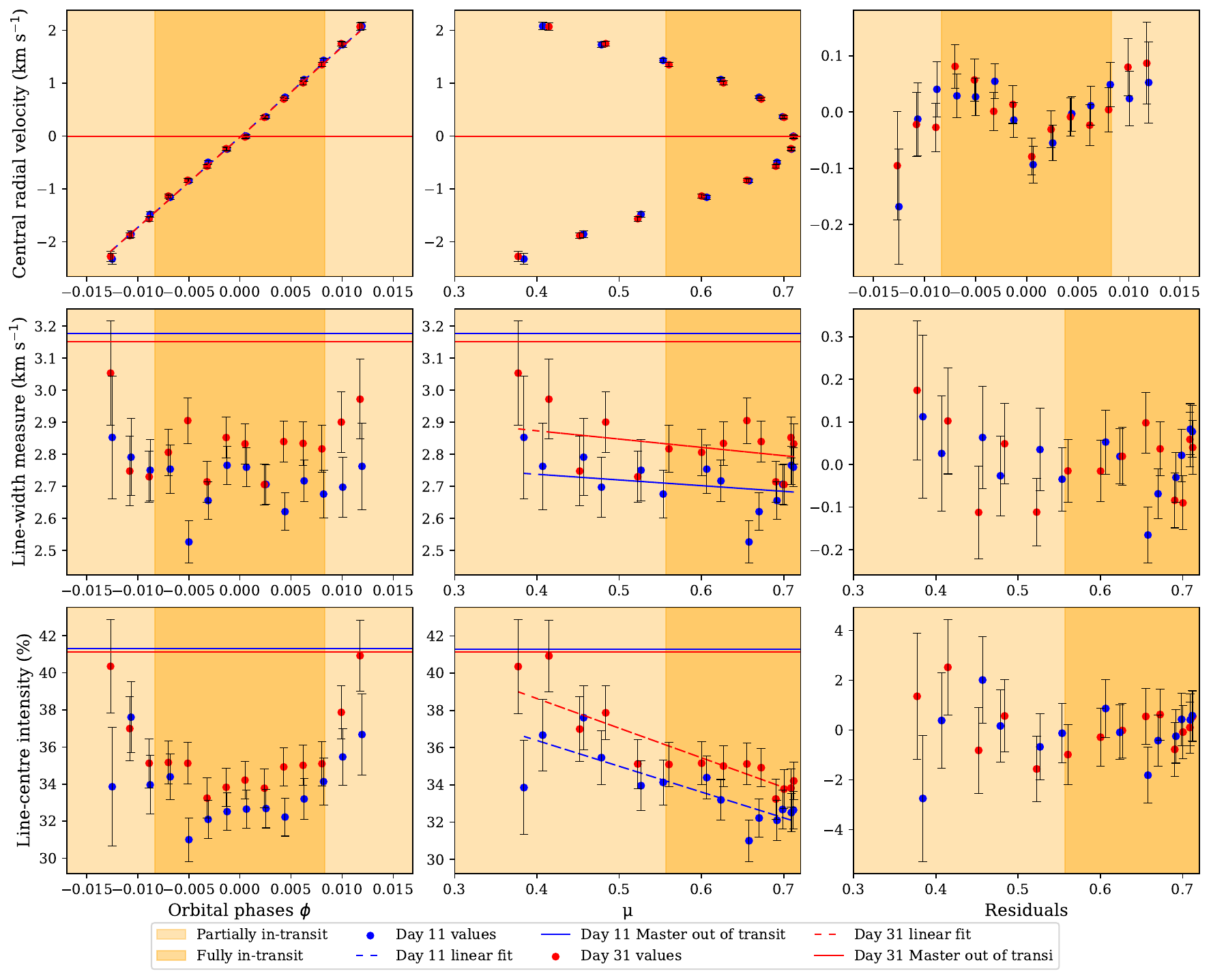}
    \caption{Same as Fig. \ref{fig:results_weakfei_ccf}, but for the strong \ion{Fe}{I} lines CCFs.}
    \label{fig:results_strongfei_ccf}
\end{figure*}

With the local \ion{Fe}{I} CCFs, we fit the profiles with the function given in Eq. \ref{eq:2.1} and extracted the three describing quantities discussed before; The central radial velocity, the line width and the line-centre intensity, shown in Figs. \ref{fig:results_weakfei_ccf} and \ref{fig:results_strongfei_ccf} for the weak \ion{Fe}{I} CCFs and the strong \ion{Fe}{I} CCFs, respectively. In these figures, the three parameters have been plotted both as a function of their respective orbital phases ($\phi$) and $\mu$, with values in the interval of $[0,1]$; however, since the planet is off-centre from the stellar disc, it reaches no higher than $\mu \approx0.72$.

At first glance, notable trends are apparent in Figs. \ref{fig:results_weakfei_ccf} and \ref{fig:results_strongfei_ccf}, most prominently the near-linear relationship between the radial velocity of the local CCFs and the orbital phase. This is an expected outcome, as we are effectively tracing the rotational motion of the stellar surface; as a first-order approximation, this is the natural consequence of a constant angular speed term in the star. However, real stars present differential rotation that is latitude-dependent, as seen in, for example, \citet{Cegla_2016}, which is more prevalent near the limb. Additionally, the radial velocity values of the disc-integrated star, derived from the master out-of-transit CCF, remain close to zero due to our prior correction to the stellar rest frame, as indicated in the continuous red and blue horizontal lines in the top left panels of Figs. \ref{fig:results_weakfei_ccf} and \ref{fig:results_strongfei_ccf}.

Further trends emerge when examining the line width and line-centre intensity, particularly for the strong \ion{Fe}{I} CCFs (middle column, centre and bottom panels of Fig \ref{fig:results_strongfei_ccf}). Both quantities exhibit a decreasing trend with increasing $\mu$, indicating a decrease in profile depth from disc centre to limb. However, this behaviour appears less consistent in the data from the 31 August compared to the 11 August, with the discrepancy most apparent in the weak \ion{Fe}{I} line measurements (middle column, centre and bottom panels of Fig. \ref{fig:results_weakfei_ccf}).

The stark difference between the two days might be due to an anomaly in the dataset, which was first noted by \citet{Cristo2024} based on the same data. As shown in Fig. \ref{fig:master_dif} and supported by the measured difference in \ion{Ca}{II} indicator levels ($0.4616 \pm 0.0002$ for the first night and $0.4479 \pm 0.0002$ for the second), both nights exhibit different activity levels, suggesting possible variations in the active regions on the stellar disc. This anomaly, which \citet{Cristo2024} identified while fitting the RM effect to the RVs from white light CCFs, appears at the end of the ingress of the transit on August 31 \citep[see Fig.~C.1 of][]{Cristo2024}. The cause for this anomaly was hypothesised to be due to a possible spot crossing event. In the subsequent work of \citet{Mounzer_2025} that used the same spectra, they found the anomaly was consistent with a spot occultation event using a spot plus transit model. Such an event could bias the data observed near ingress as we would be probing cooler regions with less convection than a quiet photosphere.

A comparison between Fig. 10 of \citet{Dravins_2018} and Figs. \ref{fig:results_weakfei_ccf} and \ref{fig:results_strongfei_ccf} highlights the improvements achieved with the new dataset.
In particular, the central radial velocity (top left panels of Figs. \ref{fig:results_weakfei_ccf} and \ref{fig:results_strongfei_ccf}) shows a clear linear correlation with orbital phase ($\phi$) with a Pearson correlation coefficient of at least 0.998 for all four datasets of local CCFs, which contrasts with the curvature observed in the results of \citet{Dravins_2018}. 
This discrepancy may stem from a combination of the lower spectral resolution of HARPS ($\lambda/\Delta\lambda \sim 115000$) compared to ESPRESSO ($\lambda/\Delta\lambda \sim 140000$). This resulted in their line profiles being instrumentally broadened and differences in the data processing methodology. To increase the S/N \citet{Dravins_2018} grouped their observations in sets of consecutive four and averaged the resulting profiles. As the planet moves along the transit chord, this temporal averaging corresponds to a spatial smearing of the occulted stellar regions, which could bias the recovered local CCFs and, consequently, the retrieved RVs.
It also explains why the absolute values of their measured line widths are higher in absolute value than those in our results, as their profiles are broader as a consequence of these two factors.

Since granulation on the stellar surface involves plasma flows in different directions, along with the uneven, corrugated structure of the granules. As we move from the centre to the limb of the stellar disc, our line of sight increasingly samples plasma flows at different angles. This causes the observed convection velocity field to become more dispersed, leading to a broadening of the line profiles. However, as stated in \citet{Dravins_2018} these centre-to-limb variations of line widths would be most visible for F-type stars; moving to colder G-type and even colder K-type stars, the stellar granulation contrast decreases with the increase in optical depth making this variation harder to detect. Indeed, no clear correlation was detected in \citet{Dravins_2018} for the line widths with $\mu$.
In our data, a weak correlation appears to be present visually (centre middle panel of Fig. \ref{fig:results_strongfei_ccf}), but we defer to a more detailed analysis in the following section where we test those correlations. Lastly, at first glance, the line-centre intensities clearly show that the \ion{Fe}{I} lines get deeper, as we move from the stellar limb to the disc centre, as theoretically predicted \citep[see Fig. 11 of][]{Dravins_2018} and discussed in the literature  \citep[see Fig. 11 of][]{Cegla_2018}

\subsection{Statistical significance of the results}

To evaluate the statistical significance of our results, regarding
the relations between intensity as well as width of the weak and strong
Fe I lines CCFs and, for the two observation epochs (epochs 11 and 31),
we performed a Bayesian model comparison by comparing the evidence
(or marginal likelihood) for several simple hypotheses. All marginal likelihoods
were estimated using nested sampling, as implemented in the Python package \texttt{Dynesty} \citep{speagle_2020,sergey_koposov_2025_17268284,skilling_2004,skilling_2006}.

First, we compared the marginal likelihoods of two distinct linear models
for such relations, assuming the slope to be zero and the other allowing
for an unconstrained (a priori), non-zero slope. Then, if for any of the four relations
considered across the two observation epochs, there was any evidence to support the model
with a non-zero slope, a further comparison was performed between
linear models with the slope constrained a priori to be either positive or negative.

A summary of our findings can be found in Table \ref{tab:bayes_factors}, in the form of natural logarithms of Bayes factors, which we simply refer to as 'Bayes factors' for simplicity. We followed the criteria of \citet{Kass95bayesfactors}, which considers that a Bayes factor greater than 10 (around 2.3 when taking the natural logarithm) provides strong evidence in favour of one model over another.

\begin{table}[h]
\centering
\caption{Bayes factors of the statistical tests on the significance of   the  profile parameters correlations with $\mu$.}
\label{tab:bayes_factors}
\renewcommand{\arraystretch}{1.3}
\begin{tabular}{lccc}
\hline
\hline
Parameter & Epoch 11 & Epoch 31 & Combined \\
\hline
Intensity ($\text{Fe}_{\text{s}}$) & 4.1/7.3 & 5.7/9.1 &6.1/9.6  \\
Width ($\text{Fe}_{\text{s}}$)     & -2.3/--  & -1.0/--  &-2.6/--\\
Intensity ($\text{Fe}_{\text{w}}$) & 21.3/25.5 & -1.3/--  & 6.4/10.5 \\
Width ($\text{Fe}_{\text{w}}$)     & 9.6/15.5  & -1.5/-- & -0.9/--  \\
\hline
\end{tabular}
\tablefoot{The table contains the natural logarithms of the Bayes factors of the statistical tests made for the inferred intensity and width of the \ion{Fe}{I} CCFs across the two observation epochs (Epoch 11 and Epoch 31) and their combination using a linear model. The first value is that of the model comparison between an unconstrained slope and zero slope, a value greater than 2.3 supports an unconstrained slope. If an unconstrained slope is supported, we give the Bayes factor between a negative slope versus a positive slope, where a value greater than 2.3 supports a negative slope; conversely, a value smaller than -2.3 supports the positive slope.}
\end{table}

We confirm that the line-centre intensity of the strong \ion{Fe}{I} CCFs decreases with increasing $\mu$, namely, the CCF profiles become less deep from the centre to the limb (bottom-middle panel of Fig. \ref{fig:results_strongfei_ccf}). This trend was observed for both individual transit epochs and for the combined data, as indicated by the Bayes factors in Table \ref{tab:bayes_factors}. The weak \ion{Fe}{I} CCFs show the same trend in the 11 August data (Bayes factor of 25.5) and in the combined dataset (Bayes factor of 10.5; see the bottom middle panel of Fig. \ref{fig:results_weakfei_ccf}). Thus, the \ion{Fe}{I} line profiles become less deep as we move from the centre of the disc to the limb, consistent with the simulations of \citet{Dravins_2018} and \citet{Cegla_2018}.

For the weak \ion{Fe}{I} CCFs on 11 August, we find that the line widths decrease with increasing $\mu$ with a Bayes factor of 15.5 (Table \ref{tab:bayes_factors}), which strongly supports this trend (i.e. the CCF profiles get wider from centre to limb). This variation is less evident than the change in line-centre intensity. In Fig. 11 of \citet{Dravins_2018}, the CLV in line widths is less prominent for K-type stars compared to G- and F-type stars, so the weaker signal observed for HD 189733 is consistent with expectations.

Finally, inspection of Figs.~\ref{fig:results_weakfei_ccf} and \ref{fig:results_strongfei_ccf} shows that in the panels corresponding to the line-width measure and line-centre intensity (see the panels in the bottom two rows for either the left or middle columns), the local CCF measurements (blue and red points) exhibit systematically lower values than those of the master out-of-transit CCF (solid blue and red lines), which represents the disc-integrated stellar profile. This behaviour is expected, as the master out-of-transit line profiles result from the combination of multiple local profiles formed across the stellar disc and Doppler-shifted by different projected rotational velocities, leading to broader and shallower integrated lines.

\begin{figure*}
    \centering
    \includegraphics[width=\linewidth]{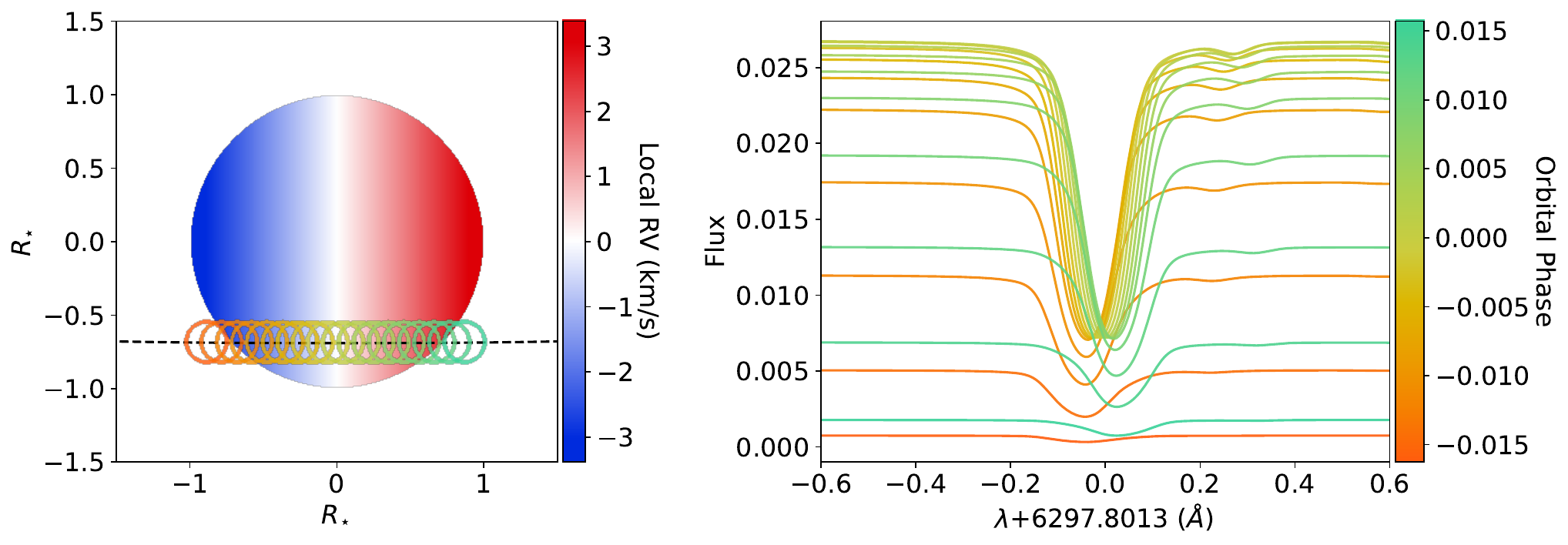}
    \caption{Orbital configuration and behind-the-planet local profiles of the 6297.8013 $\AA$ Fe$\,$I line for 11 August are shown simulated with \texttt{SOAPv4}. Left: Representation of the stellar disc in terms of velocities, along with the position of the planet as a function of the observed orbital phases. Right: Simulated behind-the-planet line profiles as a function of the orbital phase. The colours on the plot correspond to the phases and positions shown on the left plot.}
    \label{fig:shadow_simulated_11}
\end{figure*}

\subsection{Model comparison of the line properties}

The Doppler shadow technique measures local variations in stellar spectral lines as a planet occults different regions of the stellar disc during transit. By comparing theoretical predictions of local stellar spectra with observations, this method not only serves as a test for synthetic spectra, but also provides insight into (along with a means of measuring) the physical processes governing spectral line variations. These include stellar rotation, the projection of convective velocity fields, and centre-to-limb effects arising from granulation, by enabling measurements of the convective blueshift.

To model the planet-occulted spectra, we used \texttt{SOApv4} again, which is also capable of simulating time series of integrated spectra both with and without the effects of stellar activity, as well as calculating the impact of planetary transits on the integrated spectra. To compute the planet-occulted spectra, the code constructs a stellar grid that accounts for local flux variations (limb darkening), local stellar velocities due to rotation, and a grid of spectra reflecting the stellar properties. This grid can range from a single spectrum used to populate the entire grid to a matrix of limb-angle-dependent spectra that capture local variations in line profiles.

For the simulations presented here, we adopted the orbital and planetary parameters listed in Table~\ref{tab:1}. The stellar disc simulation was discretised using a $600\times600$ grid. 
An input spectrum is provided to \texttt{SOAPv4} and used as a reference local profile. Each grid point is assigned its own spectrum by Doppler shifting this profile according to the local projected rotational velocity and by adjusting its brightness through limb darkening. The local stellar spectrum is then obtained by summing the contributions of the grid points within the planet-occulted regions, allowing the simulation of local stellar spectra for specified $\mu$ values along the transit chord.
For the input spectra itself, we selected a representative example and relatively isolated line from each line list from Tables A.1 and A.2 of \citet{Dravins_2018} used to create the CCFs masks for our simulations. For the strong \ion{Fe}{I} line, we chose the one centred at $6297.8013\,\AA$ with a line-centre intensity of 40$\%$, while for the weak \ion{Fe}{I} line, we selected the line at $5651.4716\,\AA$ with a line-centre intensity of 79$\%$ \citep{Dravins_2018}. We used three different models for the input spectra of the \texttt{SOAPv4} simulations.

The first two synthetic spectra were computed with the \texttt{Turbospectrum} code for spectral synthesis\footnote{\url{https://github.com/bertrandplez/Turbospectrum_NLTE}} \citep{plez1998,plez2012,heiter2021, larsen2022, magg2022} along with \texttt{MARCS} photospheric models \citep{gustafsson2008}\footnote{\url{https://marcs.astro.uu.se}} and spectral line lists from the VALD3 database\footnote{\url{http://vald.astro.uu.se}} \citep{Ryabchikova2015}. Using the effective temperature, metallicity and surface gravity of HD 189733 in Table 1 we generated synthetic stellar spectra for the chosen representative \ion{Fe}{I} lines under LTE conditions and under non-LTE conditions. These synthetic stellar spectra were then used as input in \texttt{SOAPv4} to fill the stellar grid and generate synthetic transit observations for HD 189733 b.
For the third model \texttt{SOAPv4} we used the Institut für Astrophysik und Geophysik (IAG) ATLAS of disc-resolved observations \citep{Ellwarth_2023}. We used solar data because it provides interesting comparisons as it naturally includes the effects of granulation unlike synthetic data, while not being of the same stellar type as HD 189733.

\begin{figure}
    \centering
    \includegraphics[width=\linewidth]{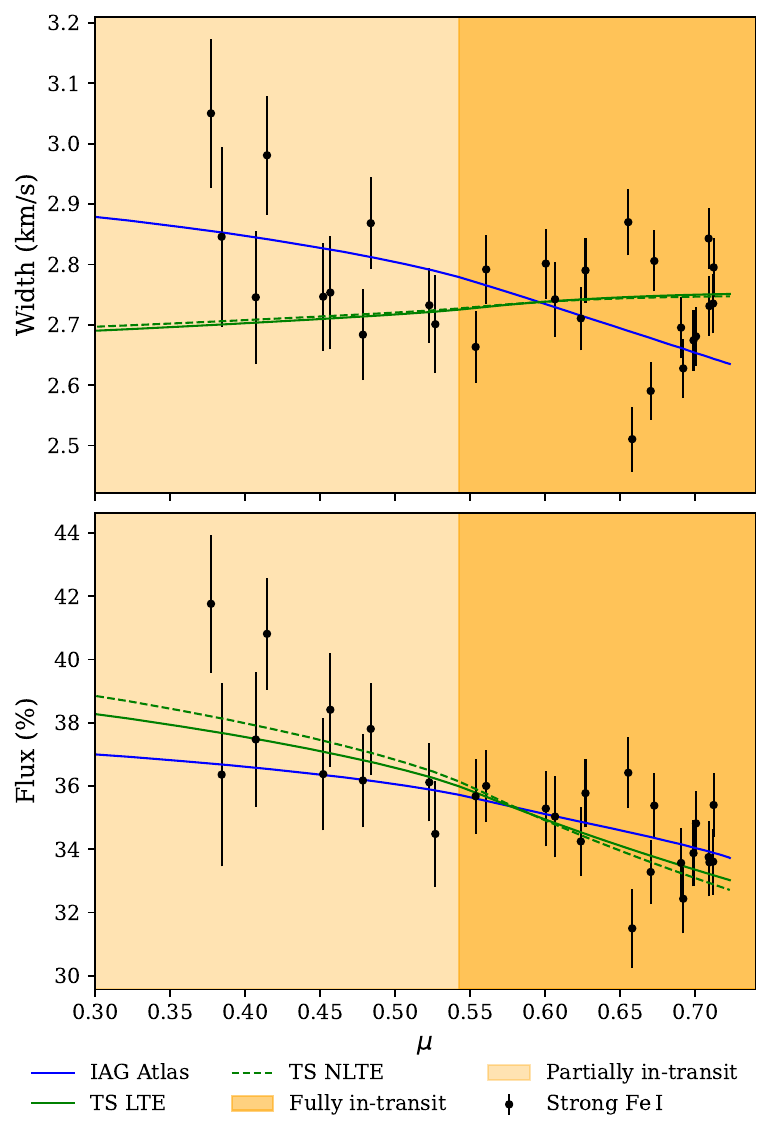}
    \caption{Width and flux relative to the continuum of the strong \ion{Fe}{I} lines are compared with the \texttt{SOAPv4} models. The line properties from the models were computed using Turbospectrum (TS in the legend) under both LTE and NLTE conditions, as well as with solar observations from the IAG ATLAS of the resolved Sun. Top: Comparison of the simulated profiles with the recovered line widths for the joint observations. Bottom: Comparison of the measured intensity flux relative to the continuum.}
    \label{fig:joint_model_SOAP_strong}
\end{figure}

The same profile quantities as the rows in Figs. \ref{fig:results_weakfei_ccf} and \ref{fig:results_strongfei_ccf} from the simulations were measured using the same modified Gaussian (Eq. \ref{eq:2.1}) as that applied to the observations. Following the same procedure as for the local \ion{Fe}{I} CCFs, we fit the profiles of the simulated planet-occulted local spectra with a modified Gaussian for both representative example lines. Figure~\ref{fig:shadow_simulated_11} illustrates the \texttt{SOAPv4} simulated local profiles for one of the representative lines and the system geometry corresponding to the observations obtained on 11 August. The resulting measurements of line width and line-centre intensity relative to the continuum are presented in Fig. \ref{fig:joint_model_SOAP_strong}.

\begin{figure}
    \centering
    \includegraphics[width=\linewidth]{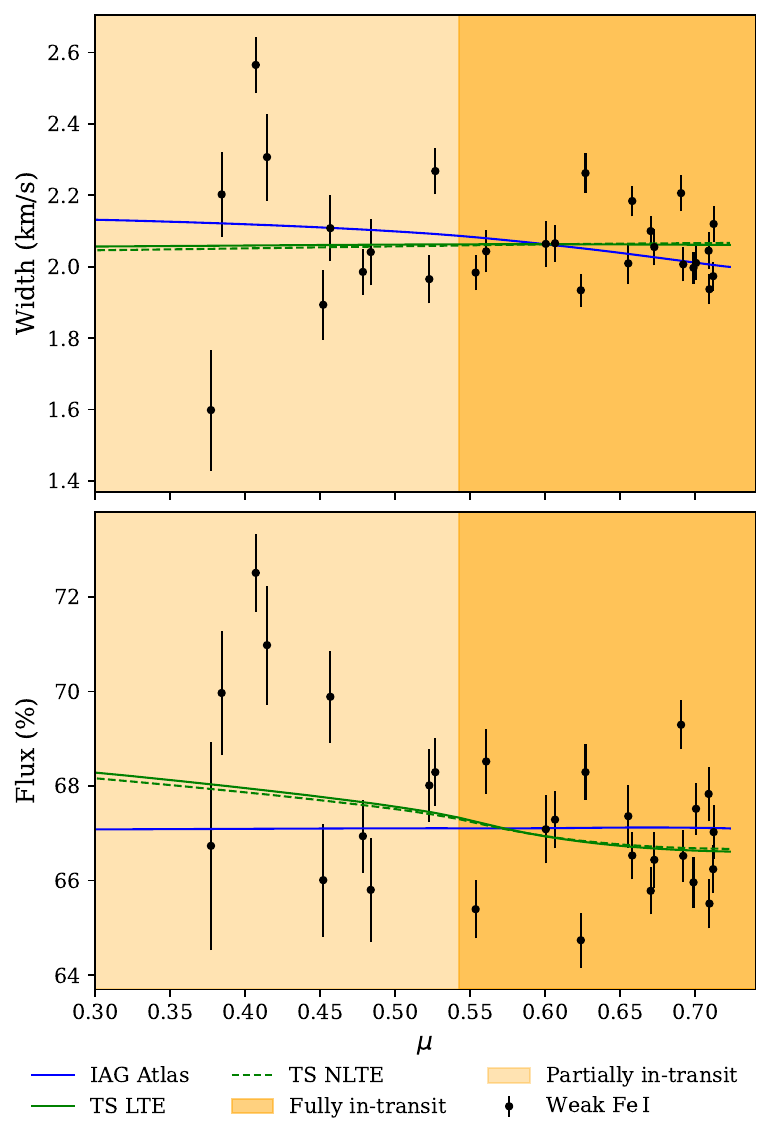}
    \caption{Width and flux relative to the continuum of the weak \ion{Fe}{I} lines are compared with the \texttt{SOAPv4} models. Same elements as Fig. \ref{fig:joint_model_SOAP_strong}.}
    \label{fig:joint_model_SOAP_weak}
\end{figure}

For the line width of the strong line in the top panel of Fig. \ref{fig:joint_model_SOAP_strong}, the \texttt{Turbospectrum} model does not exhibit significant variations under the assumptions of LTE or NLTE (similar for the weak \ion{Fe}{I} line case). These models in the same panel predict a slight increase in line width from the disc limb to the disc centre, which appears to contradict the observed data. When compared with solar simulations using observational data from the IAG ATLAS in the same panel, the model follows a similar trend but in the opposite direction. Although the simulations incorporating IAG ATLAS spectra do not correspond to the spectral type of HD 189733 (a cooler star) the agreement with the IAG ATLAS results is noticeably better. This suggests that certain physical processes may be inadequately represented in the synthetic local spectra computed with \texttt{Turbospectrum}, potentially affecting the predicted line width. A crucial missing factor in the synthetic spectra is the effect of granulation. Both the Sun and HD 189733 possess convective layers below their photospheres, and the associated granulation introduces non-thermal Doppler broadening from plasma flows and asymmetries in line profiles that are naturally present in observed spectra but absent in 1D synthetic calculations. As a result, the neglect of granulation effects in the synthetic spectra is likely to lead to underestimated line widths, contributing to the discrepancy observed here.

Regarding the line intensity of the strong line relative to the continuum in the bottom panel of Fig. \ref{fig:joint_model_SOAP_strong}, differences between the various \texttt{Turbospectrum} models (LTE vs NLTE) are more pronounced, resulting in slight variations in the predicted slopes. The simulations follow a trend similar to the observational data, although the noise level precludes a definitive distinction between the two. However, when compared with IAG ATLAS, the \texttt{Turbospectrum} simulations tend to underestimate the observed slope. This discrepancy suggests that the primary factor driving changes in spectral line intensity is the temperature difference, whereas phenomena such as granulation may have a less significant influence on line intensities.

For the weak line (Fig. \ref{fig:joint_model_SOAP_weak}), the S/N is lower due to its intrinsically lower intensity. Consequently, greater dispersion is observed in the measured quantities. The variation in line width with $\mu$ is comparable in amplitude to that computed for the strong lines. However, due to the increased dispersion, a direct comparative analysis between the models and the observational data is not feasible. Nevertheless, all models predict a relatively small variation with $\mu$, which is fully consistent with the observations. The same scenario applies to the relative flux measurements with respect to the continuum (bottom panel of Fig. \ref{fig:joint_model_SOAP_weak}). Notably, the IAG \texttt{SOAPv4} model predicts a nearly constant trend, whereas the \texttt{Turbospectrum} models indicate a slight decrease in flux from the disc limb to the disc centre, aligning with the observational data, which behaves similarly to the strong line case, meaning the profiles get deeper near the disc centre as we would expect due to less broadening of the profiles.

It is important to emphasise that the slopes presented for comparison are specific to the representative spectral lines selected in this study. Alternative line choices would alter both the intensity and width levels, potentially resulting in slight modifications to the slopes, as discussed in \citet{Dravins_2018}. Indeed, aligning the simulated profiles with the observations would require subtracting the average quantities of the simulated profiles and summing the weighted average of the observed points. This adjustment is expected, as the average CCF profile properties derived from the full line list does not directly correspond to the individual spectral lines used in the simulations.

\section{Conclusions}

The Doppler shadow method is a demonstrably useful technique for analysing centre-to-limb variations of spectral line profiles on stellar surfaces. By measuring the spectra of different regions occulted by a transiting planet, this technique can probe variations across a wide range of regions of the stellar disc with transits closer to the disc centre probing a wider range of centre-to-limb variations, and depending on the transit geometry, can also reveal latitude-dependent effects such as differential rotation. In this work, we applied the method to sets of CCFs constructed from strong and weak \ion{Fe}{I} lines from \citet{Dravins_2018} to investigate the centre-to-limb variations of the line profiles on HD 189733.

We employed a Doppler shadow methodology to analyse two transits observed with ESPRESSO, thereby extracting local CCFs of the planet-occulted regions of the stellar disc, to attempt to measure changes in the line profiles due to CLVs. From the local CCFs profiles we extracted measurements for three quantities, the central radial velocity, the line-width measure, and the line-centre intensity for both strong and weak \ion{Fe}{I} lines CCFs.

We verified a decrease in the line width from stellar limb to stellar centre for the weak \ion{Fe}{I} lines for the 11 August 2021 transit, but not for the 31 August 2021; this is possibly due to an anomaly present on the second transit. Visually for the strong \ion{Fe}{I} lines, the results match the disc-resolved data from the IAG Solar ATLAS, but do not match the \texttt{Turbospectrum} simulations (as seen in Fig. \ref{fig:joint_model_SOAP_strong}). This result suggests that some stellar atmospheric models might be limited by the absence of convective motions and granulation, which play a key role in shaping spectral line profiles. Stellar granulation produces temperature inhomogeneities and complex velocity fields in the photosphere, leading to non-thermal Doppler broadening, line asymmetries, and systematic variations of line widths from disc centre to limb that are naturally present in observed spectra, but not captured by 1D hydrostatic models. These effects have been shown to significantly impact iron line formation in stars \citep{Ausplund_2000}. In addition, granulation produces a corrugated stellar surface, such that plasma flows may be obscured or preferentially sampled along the line of sight, contributing to the observed centre-to-limb variations in line-width measure \citep{Cegla_2018}. Although HD 189733 and the Sun are of different spectral types, both possess convective envelopes, making granulation a plausible contributor to the discrepancy observed here.

In contrast, for the profile intensity of the strong \ion{Fe}{I} CCFs (the weak data set presenting a larger dispersion), the observations of HD 189733 are consistent with the Turbospectrum simulations. However, given the uncertainties and the limited set of \ion{Fe}{I} lines analysed here, no clear preference between LTE and NLTE conditions can be established. It is worth noting that other spectral lines or species, which are more sensitive to departures from LTE, might exhibit stronger differences between LTE and NLTE predictions (e.g. \citealp{amarsi_2016, Bergemann_2019}). Again, with respect to the centre-to-limb variation of the intensity of the strong \ion{Fe}{I} CCFs, a difference in the gradient of line-centre intensity as a function of $\mu$ is observed between the data for HD 189733 and the IAG ATLAS model. This behaviour is consistent with the simulated centre-to-limb intensity curves shown in Fig.~11 of \citet{Dravins_2018}. A qualitative comparison of these simulations indicates that hotter G and F type stars exhibit shallower variations in line-centre intensity with $\mu$ than cooler K-type stars, such as HD 189733. This trend may reflect the combined effects of granulation-induced temperature and velocity inhomogeneities, the depth of line formation, and projection effects, which all together influence how spectral lines respond across the stellar disc \citep{Dravins_2018}.

The results given by the Doppler shadow method provide the unique ability to test stellar line formation models allowing for a better disentanglement of stellar spectral data from planetary data when characterising exoplanets. The extension of the study of centre-to-limb variations to individual lines and different chemical species and stellar classes is thus of great importance to inform detailed transmission spectroscopy measurements and to the understanding of stellar physics. This will be the subject of future work, for instance, focussed on a direct comparison between the Sun and solar analogues.

Finally, it is worth noting that studies of centre-to-limb variations and planetary Doppler shadows will benefit from upcoming facilities. For the Sun, the PoET solar telescope \citep{Santos_2025} will provide high spectral resolution observations across the solar disc, enabling the precise mapping of line-profile variations with $\mu$. For exoplanet host stars, the ANDES spectrograph on the ELT \citep{marconi_2024, Palle2025} will offer similar spectral resolution to current high-resolution instruments such as ESPRESSO, but its much larger telescope aperture will allow higher S/Ns in comparable exposure times. This increased photon collection will improve the precision for Doppler shadow measurements during transits and enable the study of fainter or more distant stars, thereby expanding the number and diversity of targets for detailed line-profile analysis.

\begin{acknowledgements}
      Funded by the European Union (ERC, FIERCE, 101052347). Views and opinions expressed are, however, those of the author(s) only and do not necessarily reflect those of the European Union or the European Research Council. Neither the European Union nor the granting authority can be held responsible for them. This work was supported by FCT - Fundação para a Ciência e a Tecnologia through national funds by these grants: UIDB/04434/2020 DOI: 10.54499/UIDB/04434/2020, UIDP/04434/2020 DOI: 10.54499/UIDP/04434/2020.

      R.A. acknowledges the Swiss National Science Foundation (SNSF) support under the Post-Doc Mobility grant P500PT\_222212 and the support of the Institut Trottier de Recherche sur les Exoplanètes (IREx).
      This work has been carried out within the framework of the National Centre of Competence in Research PlanetS supported by the Swiss National Science Foundation. The authors acknowledge the financial support of the SNSF.
\end{acknowledgements}

\bibliographystyle{aa}
\bibliography{Bibliography}

\begin{appendix}
\nolinenumbers

\onecolumn
\section{Telluric correction}

\begin{figure}[htbp]
    \centering
    \begin{minipage}{0.48\textwidth}
        \centering
        \includegraphics[width=\textwidth]{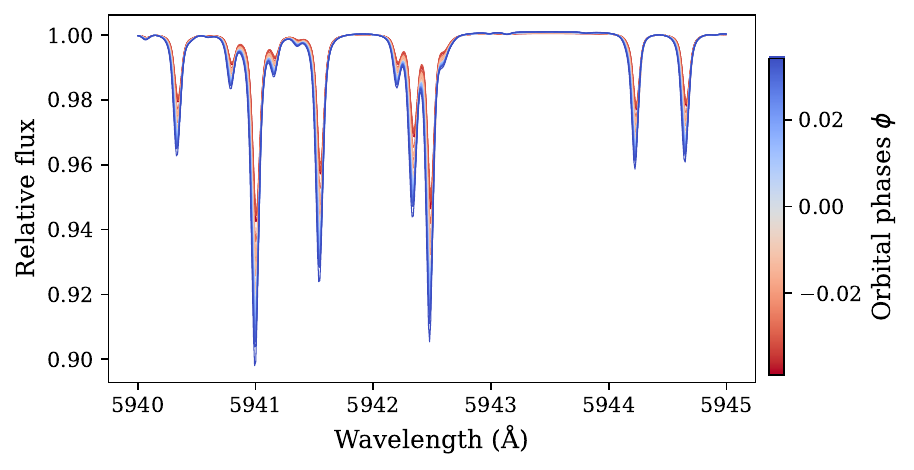}
        \includegraphics[width=\textwidth]{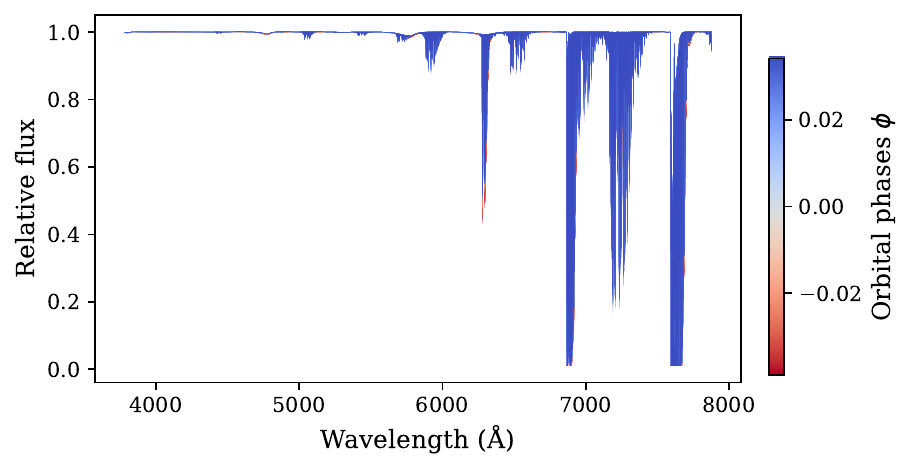}
        \includegraphics[width=\textwidth]{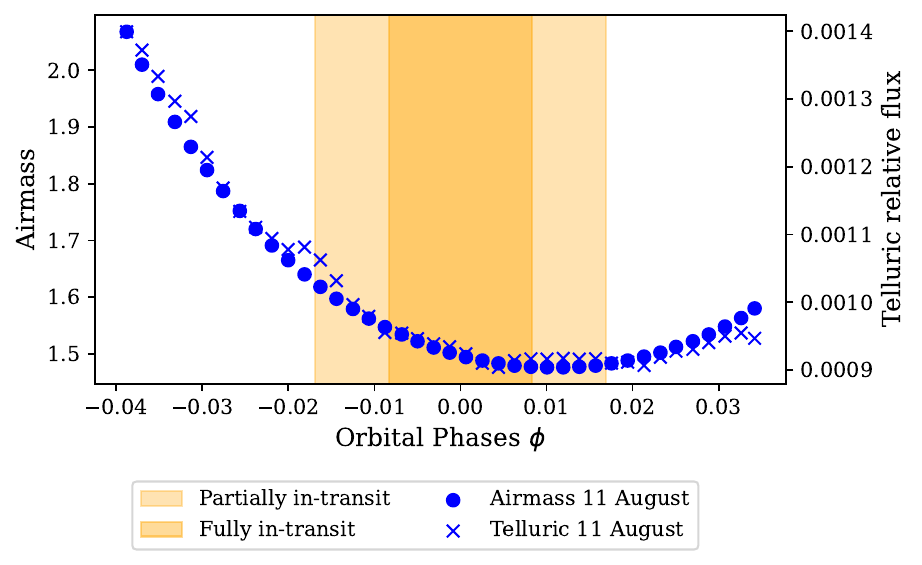}
        \caption{Top: Ratio of the uncorrected spectra over telluric-corrected spectra with \texttt{Molecfit} for 11 August 2021 with the same wavelength interval as in Figure \ref{fig:telluric}. At the middle: zoomed-out section of the same telluric spectra as the top panel, the wavelength range covers the entire domain of ESPRESSO. Colour bar indicates the orbital phase of the corresponding spectra. Bottom: Airmass and mean flux of the telluric-corrected spectra minus uncorrected spectra between 436 nm and 686 nm (range of the \ion{Fe}{I} line CCFs) as dots and crosses respectively. The shaded region is the interval of transit with the ingress and egress being lighter.}
        \label{fig:t_left}
    \end{minipage}
    \hfill
    \begin{minipage}{0.48\textwidth}
        \centering
        \includegraphics[width=\textwidth]{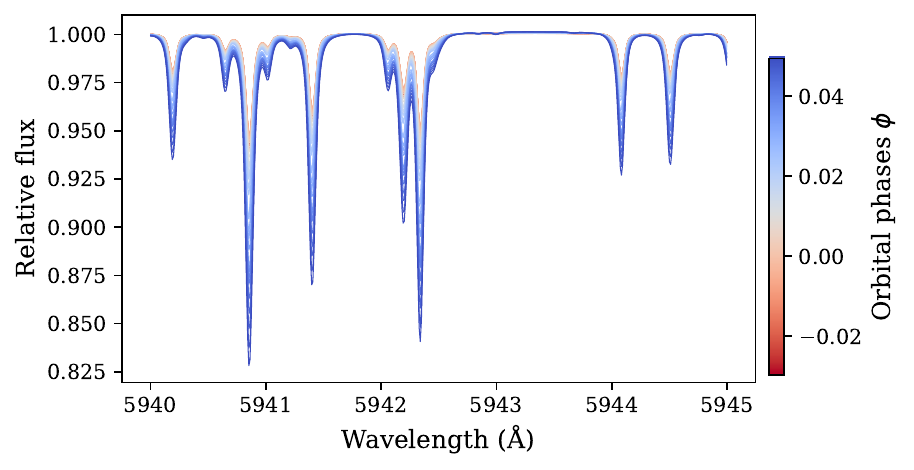}
        \includegraphics[width=\textwidth]{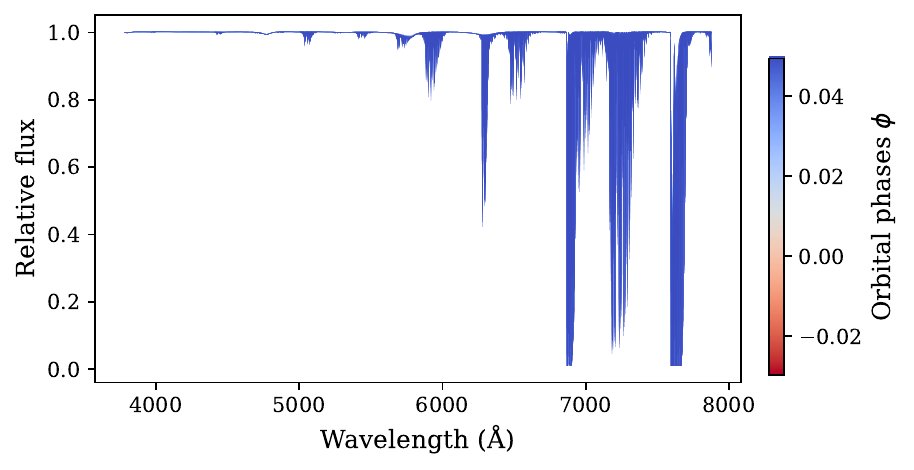}
        \includegraphics[width=\textwidth]{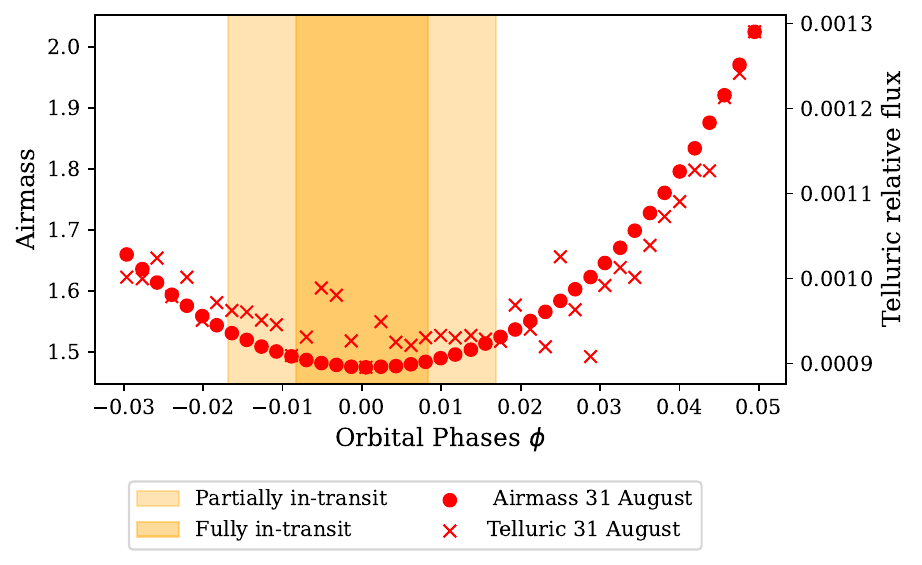}
        \caption{Top: Ratio of the uncorrected spectra over telluric-corrected spectra with \texttt{Molecfit} for 31 August 2021 with the same wavelength interval as in Figure \ref{fig:telluric}. Middle: Zoomed-out section of the same telluric spectra as the top panel, the wavelength range covers the entire domain of ESPRESSO. Colour bar indicates the orbital phase of the corresponding spectra. Bottom: Airmass and mean flux of the telluric-corrected spectra minus uncorrected spectra between 436 nm and 686 nm (range of the \ion{Fe}{I} line CCFs) as dots and crosses, respectively. The shaded region is the interval of transit with the ingress and egress being lighter.}
        \label{fig:t_right}
    \end{minipage}
\end{figure}

This section shows the variations in flux of the telluric spectra generated by \texttt{Molecfit} for telluric correction. To show the relative variation, the stellar spectra before and after correction were normalised and subtracted from each other. The average flux of the telluric correction is correlated with the airmass as we would expect from Earth's atmospheric absorption, as seen in Figs. \ref{fig:t_left} and \ref{fig:t_right}.

\clearpage

\section{Corrected RV values for the \ion{Fe}{I} lines masks}

\begin{figure*}[h!]
    \centering
    \includegraphics[width=1.\linewidth]{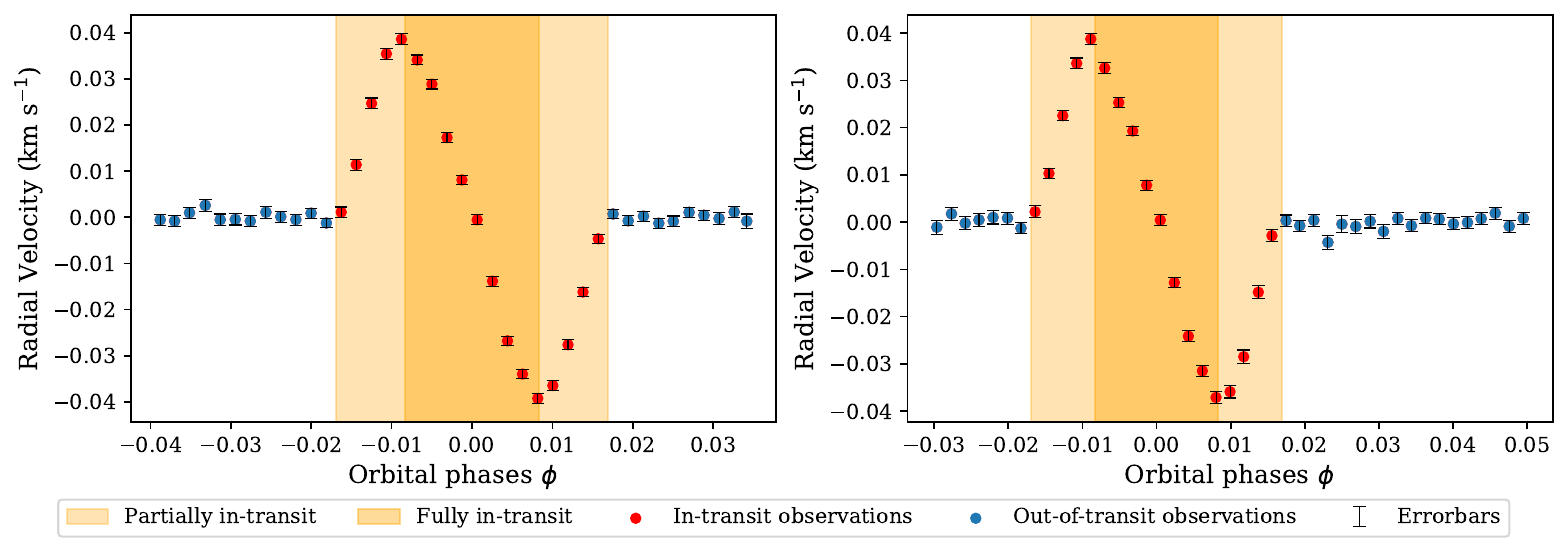}
    \caption{Barycentric-corrected RV values of the strong \ion{Fe}{I} CCFs for the two nights of observation 11 August 2021 on the left and 31 August 2021 on the right, with the same legend as in Fig. \ref{fig:bary_strongfe}. The linear fit has values $m=-1.295 \pm 0.008$ $\rm km\,s^{-1}$ and $b=-2.2255 \pm 0.0002 $ $\rm km\,s^{-1}$ for day 11 and values of $m=-1.217 \pm 0.009$ $\rm km\,s^{-1}$ and $b=-2.2365 \pm 0.0003 $ $\rm km\,s^{-1}$ for day 31.}
    \label{fig:appendix_bary_strongfe_2}
\end{figure*}

\begin{figure*}[h!]
    \centering
    \includegraphics[width=1.\linewidth]{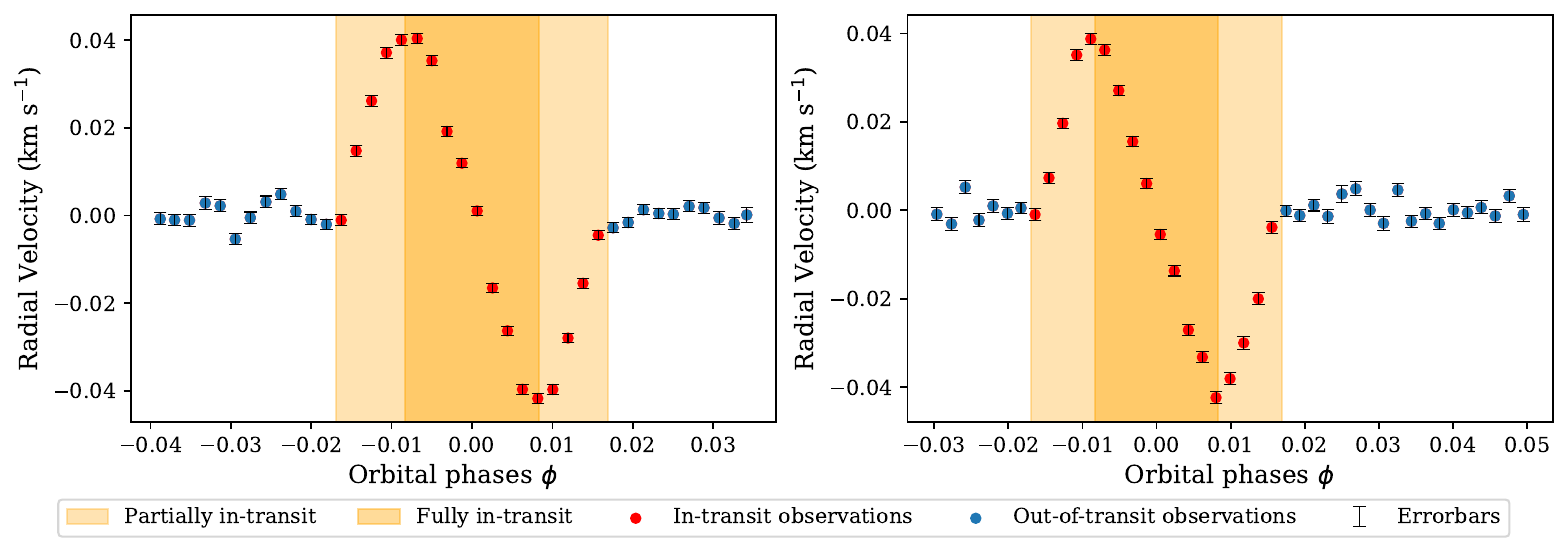}
    \caption{Barycentric-corrected RV values of the weak \ion{Fe}{I} CCFs for the two nights of observation 11 August 2021 on the left and 31 August 2021 on the right, with the same legend as in Fig. \ref{fig:bary_strongfe}. The linear fit has values $m=-1.29 \pm 0.02 $ $\rm km\,s^{-1}$ and $b=-2.0853 \pm 0.0005 $ $\rm km\,s^{-1}$ for day 11 and values of $m=-1.23 \pm 0.02 $ $\rm km\,s^{-1}$ and $b=-2.0952 \pm 0.0006 $ $\rm km\,s^{-1}$ for day 31.}
    \label{fig:appendix_bary_weakfe}
\end{figure*}

This section presents plots containing the radial velocity of each of the \ion{Fe}{I} CCFa after RV correction to the stellar rest frame showcasing the RM effect. The values of slope ($m$) and intercept ($b$) of the linear fit ($\text{RV}=m\phi+b$) are mentioned in the captions of Figs. \ref{fig:appendix_bary_strongfe_2} and \ref{fig:appendix_bary_weakfe}.

\twocolumn

\section{Uncertainties}
\label{Appendix_C}

The propagation of uncertainties is a critical step in understanding the significance of the data. Many steps were required to propagate our uncertainties from the raw spectra to our results.

The general formula for the propagation of uncertainties taking into account covariances of a function of the type $y=\text{f}(x_1,x_2,...,x_N)$ is of the form \citep{Gardner_2003}:

\begin{equation}
    \sigma_{y}^{2}=\sum_{i=1}^{N}\left(\frac{\delta f}{\delta x_{i}}\right)\sigma_{x_{i}}^{2}+\sum_{i=1}^{N}\sum_{j\neq i=1}^{N}\frac{\delta f}{\delta x_{i}}\frac{\delta f}{\delta x_{j}}\sigma_{x_{i},x_{j}}
    \label{eq:A.0}
\end{equation}

where $\sigma_{y}$ is the standard deviation of $y$, likewise for the other variables and $\sigma_{x_{i},x_{j}}$ is the covariance between $x_i$ and $x_j$.

The raw spectra contained information about the flux and its respective errors.  To propagate these uncertainties when performing cross-correlation, we started by sampling the spectra. For each data point of a spectrum, we took N samples from a Gaussian distribution with the mean being the flux value of the data point and standard deviation as the flux error, having done this for all data points of a spectrum we were left with N sampled spectra, with these sampled spectra we calculated their CCF resulting in N sampled CCFs. With the N sampled CCFs of the same original spectrum we calculated their covariance matrix which proved more useful than simple standard deviations. In total, 1000 samples were used for each CCF.

The next step was to take the CCFs and fit them with a modified Gaussian function (Eq. \ref{eq:2.1}) using $y_0$ for normalisation and $x_0$ for barycentric correction. For the fitting we used the curve\_fit function from the Python package \texttt{scipy}, which accepts uncertainties as weights and provides the covariance matrix of our fitting parameters, taking the square root of the diagonals we obtain the standard deviations of the fitting parameters that we use as uncertainties.

When normalising the CCFs by $y_0$ ($f=\frac{F}{y_0}$), ignoring covariances, the uncertainties are given via

\begin{equation}
    \sigma_{f}=f\sqrt{\left(\frac{\sigma_{y_{0}}}{y_{0}}\right)^{2}+\left(\frac{\sigma_{F}}{F}\right)^{2}}
    \label{eq:A.1}
,\end{equation}

where $f$ is the normalised flux, $F$ is the flux of the CCF prior to normalisation, and $\sigma$ represents the uncertainties.

However, since we have the covariances of the CCFs $\left(f_{i}=\frac{F_{i}}{y_0}\right)$, the propagation was carried out with the following formula derived using Eq. \ref{eq:A.0}:

\begin{equation}
    \sigma_{f_{i},f_{j}}=\begin{cases}
        \frac{F_{i}F_{j}}{y_{0}^{2}}\left(\left(\frac{\sigma_{y_{0}}}{y_{0}}\right)^{2}+\frac{\sigma_{F_{i},F_{j}}}{F_{i}F_{j}}\right)+\frac{F_{i}F_{j}}{y_{0}^{2}} & i\neq j\\
        \frac{F_{i}F_{j}}{y_{0}^{2}}\left(\left(\frac{\sigma_{y_{0}}}{y_{0}}\right)^{2}+\frac{\sigma_{F_{i},F_{j}}}{F_{i}F_{j}}\right) & i=j
        \end{cases}
    \label{eq:A.2}
\end{equation}

Here, $\sigma_{F_{i},F_{j}}$ is the covariance between the $i$th and $j$th CCF flux values. This formula is similar to \ref{eq:A.1} and simplifies for the $i=j$ case. The extra term comes from the symmetry of the covariance matrix.

The RV correction is then performed to place the CCFs in the rest frame of the star. The next step was the interpolation of the CCFs to share the same RV axis. This was done to calculate the master out-of-transit and perform the subtraction. We used a simple linear interpolation. However, propagating the uncertainties proved to be more complicated. At first the following formula was used:

\begin{equation}
    \sigma_{f_{i}}=\sqrt{\left(\frac{rv_{i-1}-rv_{i}}{rv_{i+1}-rv_{i-1}}+1\right)^{2}\sigma_{f_{i-1}}^{2}+\left(\frac{rv_{i}-rv_{i-1}}{rv_{i+1}-rv_{i-1}}\right)^{2}\sigma_{f_{i+1}}^{2}}
    \label{eq:A.4}
\end{equation}

where $rv$ represents the radial velocities of the respective flux values for each of the CCF points. The index, $i$, represents the new interpolated point for the chosen $rv_i$ value for the new $x$-axis, the $i-1$ and $i+1$ refer to points of the original CCF before and after the interpolated point.

\begin{figure}[h!]
    \centering
    \subfloat{\includegraphics[width=.95\linewidth]{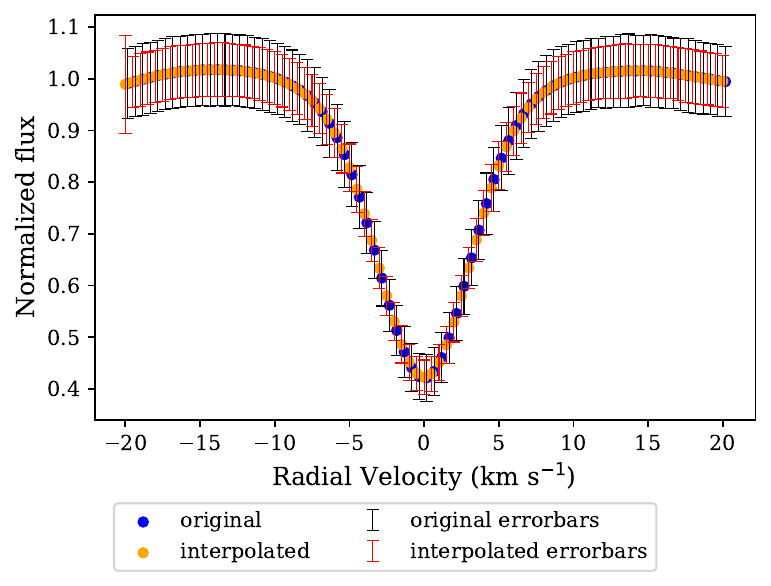}}\\
    \subfloat{\includegraphics[width=.95\linewidth]{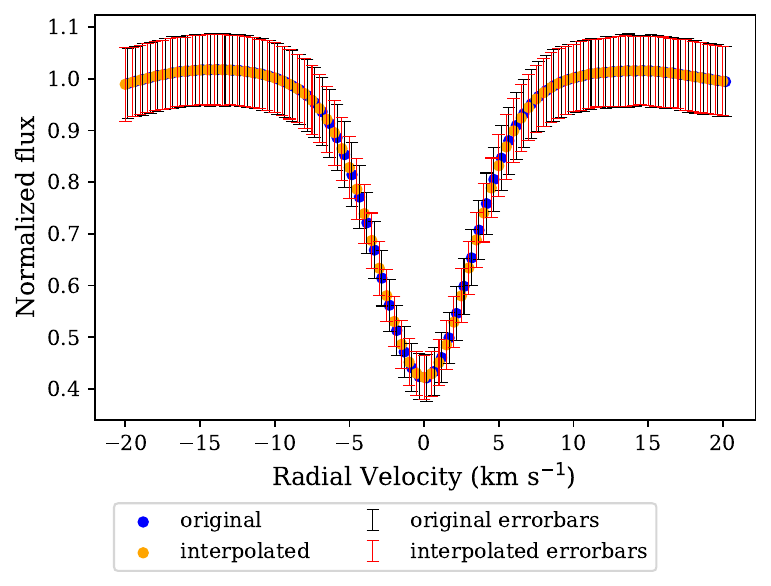}}
    \caption{Top: interpolation of a strong \ion{Fe}{I} CCF without taking into account uncertainties using formula \ref{eq:A.4}. Bottom: same as the top except we consider covariances between points using formula \ref{eq:A.5}. The blue points are the CCF prior to interpolation, while orange is after interpolation. The black error bars are the uncertainties (square root of the diagonal of the covariance matrix) of the CCFs before interpolation with the red error bars being after interpolation. The uncertainties have been scaled up by a factor of 100 for ease of viewing.}
    \label{fig:inter_comp}
  \end{figure}

However, despite the simplicity of linear interpolation, this formula is an oversimplification that assumes no correlation between points. As such, the uncertainties end up slightly underestimated, as seen in the comparison in Fig. \ref{fig:inter_comp}. We found that a better propagation could be achieved if we took covariances into account, which for our data could only be obtained from sampling the spectra and calculating the covariances of the CCFs of the samples. In the covariance matrix of the \ion{Fe}{I} CCFs there are non-diagonal terms that were found to be on the same order of the diagonal variances (around half) and by including them in the error propagation resulted in non-underestimated uncertainties. Including the non-diagonal terms gives the following formula:

\begin{equation}
    \begin{split}
        \sigma_{f_{i}}^2=\left(\frac{rv_{i-1}-rv_{i}}{rv_{i+1}-rv_{i-1}}+1\right)^{2}\sigma_{f_{i-1},f_{i-1}}+\left(\frac{rv_{i}-rv_{i-1}}{rv_{i+1}-rv_{i-1}}\right)^{2}\sigma_{f_{i+1},f_{i+1}} \\
        +2\left(\frac{rv_{i}-rv_{i-1}}{rv_{i+1}-rv_{i-1}}-\left(\frac{rv_{i-1}-rv_{i}}{rv_{i+1}-rv_{i-1}}\right)^{2}\right)\sigma_{f_{i-1},f_{i+1}}
    \end{split}
    \label{eq:A.5}
\end{equation}

The following step was the averaging of the out-of-transit CCFs to create the master out-of-transit CCF. This was straightforward due to the prior interpolation. Since we took the mean of the respective CCFs and their uncertainties are not exactly equal, we propagated the formula of the mean:

\begin{equation}
    \sigma_{f_{master,j}}=\frac{1}{M}\sqrt{\sum_{i=0}^{N-1}\sigma_{f_{i,j}}^{2}}
    \label{eq:A.6}
\end{equation}

where the index $j$ refers to the $j$th point on a given CCF, while the index $i$ refers to a specific out-of-transit CCF with $M$ being the total amount of out-of-transit CCFs.

Finally, we get to the subtraction to arrive at the local CCFs. This was done using formula \ref{eq:2.sub}, propagating yields the following relation:

\begin{equation}
    \sigma_{F_{\rm local}}=\sqrt{\sigma_{f_{\rm Master,out}}^{2}+\left(\sigma_{f_{\rm in}}F_{\text{SOAP}}\right)^{2}}
    \label{eq:A.7}
\end{equation}

Next, the local CCFs representative of specific portions of the stellar surface were fitted with a modified Gaussian \ref{eq:2.1} enabling us to extract the three descriptive parameters of their profiles. These were the central radial velocity $x_0$, the line-width measures $b$ and line-centre intensities defined in Eq. \ref{eq:sus}. The first two have their uncertainties taken directly from the respective diagonal in the covariance matrix given by the scipy curve\_fit function, the propagation for the last one was taken as:

\begin{equation}
    \sigma_{I}=\frac{a}{y_{0}}\sqrt{\left(\frac{\sigma_{a}}{a}\right)^{2}+\left(\frac{\sigma_{y_{0}}}{y_{0}}\right)^{2}}\times100
    \label{eq:A.8}
\end{equation}

This concludes the propagation of uncertainties for the CCFs.

\end{appendix}
\end{document}